\documentclass[11pt]{article}
\usepackage{authblk}
\usepackage{epsf}
\usepackage{graphicx}
\usepackage{a4}
\usepackage{amsmath}
\usepackage{amssymb}
\usepackage{cite}		
\usepackage{color}
\usepackage{verbatim} 
\usepackage{empheq} 
\usepackage{tikz,pgf,pgfplots}
\usetikzlibrary{decorations.pathreplacing}

\usepackage{geometry} 
\usepackage{bbold} 

\def\titlepage{\@restonecolfalse\if@twocolumn\@restonecoltrue\onecolumn
     \else \newpage \fi \thispa gestyle{empty}\c@page\z@
        \def\thefootnote{\fnsymbol{footnote}}
	\setcounter{page}{0} }
\def\endtitlepage{\if@restonecol\twocolumn \else  \fi
        \def\thefootnote{\arabic{footnote}}
        \setcounter{fowotnote}{0}}  
        
\definecolor{c1}{rgb}{1, 0, 0}
\definecolor{c2}{rgb}{0, 1, 0}
\definecolor{c3}{rgb}{0, 0, 1}
\definecolor{c4}{rgb}{1, 0, 1}
\definecolor{c5}{rgb}{0, 1, 1}

\def\ie{\hbox{\it i.e.}}
\def\nn{\nonumber}
\def\beq{\begin{equation}}
\def\eeq{\end{equation}}
\def\bea{\begin{eqnarray}}
\def\eea{\end{eqnarray}}
\def\EQ{\begin{equation}}
\def\EN{\end{equation}}

\allowdisplaybreaks

\begin{document}
\title{On the CFT describing the spin clusters in $2d$ Potts model}

\author{Marco Picco$^1$ and Raoul Santachiara$^2$}
\affil{\small$^1$\textit{Sorbonne Universit\'e, CNRS UMR 7589, Laboratoire de Physique Th\'eorique et Hautes Energies, 
4 Place Jussieu, 75252 Paris Cedex 05, France}}
\affil{\small$^2$\textit{LPTMS, CNRS (UMR 8626), Universit\'e Paris-Saclay, 91405 Orsay, France}}
\date{\today}

\maketitle

\abstract{We have considered clusters of like spin in the $Q$-Potts model. Using  Monte Carlo simulations,
 we studied the S clusters on a toric $L_h\times L_v$ square lattice for values of $Q\in [1,4]$. We continue the work initiated in \cite{Delfino13} by measuring the universal finite size corrections of the two-point connectivity.  The numerical data are perfectly compatible with the CFT prediction, thus supporting the existence of a consistent CFT, still unknown, describing the connectivity Potts spin clusters. We provided in particular new insights on the energy field of such theory. For $Q=2$, we found a good agreement with the prediction that the Ising spin clusters behave as the Fortuin-Kasteleyn ones at the tri-critical point of the dilute $1$-Potts model. We show that the structure constants are likely to be given by the imaginary Liouville structure constants, consistently with the results in \cite{Delfino13,Ang21}. For $Q\neq 2$ instead,  the structure constants we measure do not correspond to any known bootstrap solutions. The validity of our analysis is backed up by the measures of the spin Potts cluster wrapping probability for $Q=3$. We evaluate the main critical exponents and the correction to the scaling. A new exact and compact expression for the torus one-point of the $Q-$ Potts energy field is also given.}

\newpage

\section{Introduction}
Percolation theory, that is of paramount importance in statistical and mathematical physics, studies the clusters formed by randomly activated sites or bonds of some lattice.
Particularly interesting are the percolation models where the activation probability is related to the equilibrium configurations of a statistical model. When the model undergoes a continuous phase transition in correspondence of which the clusters percolate, these latter not only provide a geometrical characterization of the critical point but also unveil new universality classes besides the one describing the critical fluctuations of the local degrees of freedom. The description of critical phenomena in terms of their percolative properties dates back to the droplet model introduced in \cite{Fisher67}.

A relevant example, studied here, is the two-dimensional Ising model whose percolative universal behaviors are encoded in two types of clusters. One is the spin (S) clusters that are formed by clusters of like spins. The other are the Fortuin-Kasteleyn (FK) clusters \cite{fk69,fk72}, that are obtained by launching a Bernoulli percolation on each S cluster, see Eq.~(\ref{FKSC}) for a precise definition. Both S and FK clusters undergo a percolation transition at the thermal Ising critical point \cite{ConiglioNappi}, at which they become conformally invariant fractals. 
The question of whether a conformal field theory (CFT) could capture the behavior of these fractals has challenged theoretical and mathematical physics for more than thirty years.  

The S and the FK clusters percolation transitions are described by a pair of two critical points, referred to as S and the FK points, connected by a renormalization group (RG) flow\cite{Coniglio80}. Such scenario extends to the S and FK clusters of the $Q$-Potts models \cite{CP, Vanderzande92} that contain the Ising case as a special case for $Q=2$. 

The FK point critical exponents  were computed long time ago by Coulomb gas techniques \cite{Nienhuis_CG}. Recently, compelling predictions on the FK multi-point connectivities came from a numerical and analytical conformal bootstrap approach\cite{devi11,Picco16,Picco19,Yifei20,Nivesvivat21}. Many structure constants upon which the CFT is built have been determined in this way. These results are unveiling the fine structure of the CFT describing the FK point (FK-CFT), paving the way for a deeper comprehension of the (logarithmic) Virasoro representations involved in these theories \cite{Nivesvivat21,Marco08,Raoul13,Gorbenko2020}.    

The S point and the CFT describing it (S-CFT) are much less known. In this respect, the Ising ($Q=2$) S point is special because it can be identified with the tri-critical point of the dilute $Q=1$ Potts model \cite{SV}. 
Moreover, the S clusters interfaces become in the continuum limit the loops of the CLE$_{3}$ loop ensemble \cite{Sheffield12}. This opens the way to the powerful tool box of probability theory. In particular the exact scaling limit of the S cluster three-point connectivity, that was proposed for the first time in \cite{Delfino13}, was recently proven by these methods \cite{Ang21}. For $Q\neq 2$, even the existence of a consistent S-CFT theory, providing for instance the S connectivity properties, remain an open question. In \cite{Delfino13}, by numerically studying the S three-point connectivity, it was shown that, if such theory exists, it would not correspond, for $Q\neq 2$, to any known CFT.

In this paper we study different aspects of the S cluster percolation transition, especially the torus two-point connectivity and the wrapping probability. We show that the predictions of a local CFT well describes the Monte Carlo numerical findings. We provide therefore  new numerical observations that support, for $Q\in[1,4]$, the existence of a consistent S-CFT. In particular we obtain new insights of its energy field and of its correlation functions.

\section{CFT basic concepts: notations, singular states and Liouville structure constants}

Let us briefly recall the CFT notations and the basic concepts we will use in this paper. We refer the reader to \cite{ribaultlect} for an exhaustive review. 

We consider CFTs with central charge $c\in [0,1]$, conveniently expressed in the $\beta$ parametrization :
\begin{align}
c = 1-6\left(\beta -\frac{1}{\beta}\right)^2, \quad \beta \in \left[\sqrt{\frac23},\sqrt{\frac32}\right]\; .
\label{betaparam}
\end{align} 
A Virasoro highest-weight representation  $\mathcal{R}_{\Delta}$ is labeled by the conformal dimension $\Delta$ of its highest-weight state. The dimension $\Delta$ is parametrized as: 
\begin{equation}
\label{paradelta}
 \Delta= \Delta_{r,s}  = \frac{c-1}{24} + \frac14 \left(r\beta -\frac{s}{\beta}\right)^2, \quad r,s \in \mathbb{R} \; .
\end{equation}
To each $c\in [0,1)$  correspond two values of $\beta$, with $\beta^2\in [2/3,1)$ and $\beta^2\in (1,3/2]$. In the following, when using the notation (\ref{paradelta}), we will assume $\beta^2\in [2/3,1]$. Notice that, under  $\beta^2\to \beta^{-2}$, one has  $\Delta_{r,s}\to\Delta_{s,r}$.    
The parametrization in Eq.~(\ref{paradelta}) is reminiscent of the minimal model $M_{p,q}$ solutions. These CFTs are defined for $\beta^2=p/q$, with $p,q$ coprime natural numbers, and contain only the Virasoro representations with $r=1, \cdots, p-1$ and $s=1,\cdots, q-1$. Note that the CFTs we are interested in are not minimal models.

We denote the spinless primary fields as $V_{\Delta}$, where $ \Delta$ is the left and right conformal dimension. 
The structure constants are given by the three-point correlation functions 
\begin{equation}
C_{\Delta_1,\Delta_2}^{\Delta_3}=\left< V_{\Delta_1}(0)V_{\Delta_2}(1)V_{\Delta_3}(\infty)\right>, 
\end{equation} 
where $\left<\cdots\right>$ is the correlation function in the CFT under consideration. We use equivalently the short-end notations  
\begin{align}
V_{\Delta_{(r,s)}}=V_{r,s}, \quad C_{(r_1,s_1),(r_2,s_2)}^{(r,s)}=\left<V_{r_1,s_1}(0)V_{r_2,s_2}(1)V_{r,s}(\infty)\right> \; .
  \end{align}
When $r,s$ are positive integers, $r,s \in \mathbb{N^*}$, the representation $\mathcal{R}_{\Delta_{r,s}}$  has a state $\chi_{rs}$ of dimension $\Delta+ r s$ with 
vanishing norm. The $\mathcal{R}_{r,s}$ is said to be degenerate al level $r \times s$. When $\chi_{rs}=0$, the fusions rules between  $V_{r,s}$ and the other fields 
are strongly constrained. This is for instance the case for the minimal models whose degenerate representations have vanishing singular states. On the other hand, 
when the theory is not unitary, the singular state does not necessarily vanish. A relevant exemple is provided by the FK-CFT. In this theory the energy field, 
$\varepsilon^{FK}$, of dimension  $\Delta_{1,2}$ has a vanishing singular state while the singular state in the representation $\mathcal{R}_{2,1}$ does not.

From the above discussion it is clear that, to probe the fine structure of a representation, it is not sufficient to know its conformal dimension, 
or equivalently the two-point functions on the plane. One needs to study quantities that involve its three points functions.

As shown in a series of papers, see \cite{Yifei20,Delfino13} and references therein,  an important role is played in the study of the FK and S clusters by the imaginary DOZZ structure constants $\left(C^{c\leq 1}\right)_{\Delta_1,\Delta_2}^{\Delta_3} $  \cite{Kostov06,Schomerus06,Zamolodchikov05}.
They are given by:  
\begin{equation}
\label{Liouc}
\left(C^{c\leq 1}\right)_{\Delta_1,\Delta_2}^{\Delta_3} = A(\beta)\frac{\Upsilon_\beta(\alpha_1+\alpha_2+\alpha_3+2\beta-1/\beta)\prod_{\sigma\in\mathcal{S}_3}\Upsilon_\beta(\alpha_{\sigma(1)}+\alpha_{\sigma(2)}-\alpha_{\sigma(3)}+\beta)}{\sqrt{\displaystyle\prod_{j=1}^3\Upsilon_\beta(2\alpha_j+\beta)\Upsilon_\beta(2\alpha_j+2\beta-1/\beta)}}
\end{equation}
where $\sigma\in \mathcal{S}_3$ indicates a permutation of the three indexes 1,2,3, the charges $\alpha$ are related to the dimensions $\Delta$ by $\Delta = \alpha(\alpha-\beta+\beta^{-1})$, and
\begin{equation}
A(\beta) = \frac{\beta^{\beta^{-2}-\beta^2-1}}{\Upsilon_\beta(\beta)}\sqrt{\gamma(\beta^2)\gamma(\beta^{-2}-1)},\quad \gamma(x) = \frac{\Gamma(x)}{\Gamma(1-x)}.
\end{equation}
The special function $\Upsilon_\beta$ has integral representation
\begin{equation}
\label{Zam_int}
\log\Upsilon_{\beta}(x)=\int_{0}^{\infty}\frac{dt}{t}\left[\frac{(Q/2-x)^2}{e^t}-\frac{\sinh^2\frac{t}{2}(Q/2-x)}{\sinh\frac{\beta t}{2}
\sinh\frac{t}{2\beta}}\right],
\end{equation}
convergent for $0<x<\beta+\beta^{-1}$. The analytic continuation for general value of $x$ can be obtained by using the shift equations,

\begin{align}
&\Upsilon_\beta(x+\beta) = \Upsilon_\beta(x)\beta^{1-2\beta x}\gamma(\beta x), \;\Upsilon_\beta(x+\frac{1}{\beta}) = \Upsilon_\beta(x)\beta^{2\frac{x}{\beta}-1}\gamma(\frac{x}{\beta})\nonumber \\
&\Upsilon_\beta(\beta+\frac{1}{\beta}-x) = \Upsilon_\beta(x).
\end{align}

The $ C^{c\leq 1}$ appear also in the description in the CLE$_{4\beta^{2}}$ loop models. For instance, the value
$\left(C^{c\leq 1}\right)^{(1/2,0)}_{(1/2,0),(1/2,0)}$, for $2/3\leq \beta^2\leq 1$, was rigourously proven to describe the three point connectivity of the loops interiors\cite{Ang21}. The Ising spin interfaces are described in the continuum limit by the CLE$_{3}$ loop model and 
$\left(C^{c\leq 1}\right)^{(1/2,0)}_{(1/2,0),(1/2,0)}$ gives for $\beta^2=3/4$ the S three-point connectivity, as first proposed and numerically verified in \cite{Delfino13}.
        
\section{The FK and S fixed points in the $Q-$Potts model}
\label{secFK}
The $Q-$Potts model is defined by the Hamiltonian:
\begin{equation}
{\cal H}_{Q}=-J\sum_{<x,y>}\delta_{s(x),s(y)}\,,\quad s(x)=1,\ldots,Q\,
\label{def:Potts}
\end{equation}
where $<x,y>$ are neighbouring sites on a, say, square lattice, and $s(x)$ is a $Q-$state spin sitting at the site $x$, $s(x)=1,\cdots Q$. For a critical value $J_c=\log(1+ \sqrt{Q})$ of the coupling, the model exhibits a ferromagnetic phase transition, 
which is of second order for $Q\leq 4$ and of first order for $Q>4$\cite{Baxter82}. In this paper we always consider: 
\begin{equation}
Q\in [1,4],
\end{equation} 
which correspond to the interval values of  $Q$ that we can simulate with our Monte Carlo alogorithm, see Section~(\ref{Meas}).
A renormalization group (RG) calculation captures the changeover from second to first order only if vacant sites are allowed \cite{Nienhuis79}. The thermodynamic behavior of the Potts model is therefore better understood if one considers its dilute version, where the configuration energy becomes~:
\begin{align}
{\cal H}_{DQ}=&-\sum_{<x,y>}n(x) n(y)\left(F+J\delta_{s(x),s(y)}\right)+\mu \sum_{x} n(x)
\label{def:dilPotts}
\end{align}
where $n(x)=0,1$ are the vacancies degree of freedom, $F$ is their lattice gas coupling and $\mu$ determines the vacancy concentration. For $\mu=-\infty$ one recovers the model in Eq.~(\ref{def:Potts}). Studying the RG flows in the parameter plane $(J,\mu)$ of the dilute Potts, one finds two critical points that merge at $Q=4$: one describes the pure critical Potts model, and the other, separating three phases, the tri-critical point\cite{Nienhuis79}. The universal critical partition functions on a torus, $Z_Q$ and $Z^{\text{tri}}_Q$, were computed in \cite{FraSaZu87}. As a consequence, the central charge of the two points, for each value of  $Q\in [1,4]$, are given by setting in Eq.~(\ref{betaparam}) the values of $\beta$  that solve the equation:  
\begin{align}
Q = 2+2\cos\left[2\pi(1- \beta^2)\right] \quad \text{with} \begin{cases} \tfrac23 \leq  \beta^2 \leq 1\ &\quad \text{Critical $Q$-Potts}\\
1 \leq  \beta^2 \leq \frac32  &\quad \text{Tri-critical $Q$-Potts}
\end{cases}.
\label{cq}
\end{align} 
The two branches of solution determine the central charge of respectively the critical and the tri-critical $Q$-Potts.
For instance, the critical $3$-Potts point has central charge $c=4/5$, corresponding to the $ \beta=\sqrt{5/6}$ solution 
of Eq.~(\ref{cq}) while the tri-critical $3$-Potts point, of central charge $c=6/7$, corresponds to the $\beta=\sqrt{7/6}$ solution branch. 

Let us focus now on the pure $Q$-Potts model and its $p_B$ clusters. The $p_B$ clusters are obtained by connecting neighboring like spins with probability $p_B$. To study these clusters, one considers the  Hamiltonian \cite{CP, Vanderzande92, Deng2004}
\beq
\label{hamDP}
\mathcal{H}_{QR}=\mathcal{H}_{Q}-K\sum_{\langle x,y\rangle}\delta_{s(x),s(y)}\left(\delta_{\tau(x),\tau(y)}-1\right)\,\quad\tau(x)=1,\dots,R
\eeq
where $K$ couples the $Q$-Potts model with an auxiliary $R$-state Potts model with spins $\tau(x)$. The value of $K$ fixes the probability $p_B$ as:
\begin{equation}
\label{p_BK}
p_B= 1-e^{-K}\;:\quad \text{$p_B$ clusters}.
\end{equation}

The percolative generating function of the $p_B$ cluster is given by $\left(d \mathcal{F}_R/d R\right)_{R=1}$, where $\mathcal{F}_R$ is the free energy,  
$e^{-\mathcal{F}_R}=\sum_{\{s,\tau\}}e^{-\mathcal{H}_{QR}}$.
The FK and S clusters are found by taking respectively: 
\begin{align}
K&=J,\; p^{\text{FK}}_B =  1-e^{-J}\;:\quad \text{FK clusters} \nonumber \\
K&=\infty, \;p^{\text{S}}_B = 1\;:\quad \text{S clusters} \; .
\label{FKSC}
\end{align}
For $K=J$, one finds back the Eq.~(\ref{def:Potts}) in the limit $R\to 1$, which explains the fact that the FK exponents coincide with the ones describing the ferromagnetic Potts transition \cite{CP}. One can analogously define the tri-critical FK clusters, whose percolation exponents coincide with the thermal and magnetic exponents of the tri-critical Potts point. For instance, the scaling exponents of the average size of the FK and of the tri-critical FK clusters coincide with the magnetic exponents of the critical and of the tri-critical Potts model, given in the Eq.~(\ref{connfk}). 

\begin{figure}
 \begin{tikzpicture}[scale = 0.7,  every node/.style={scale=0.7}]
 \draw[thick, ->] (-6, 0) -- (10,0) node[below]{$p_B = 1-e^{-K}$};
 \draw[thick] (8,0.1)--(8,-0.1);
 \draw (8,0) node[below]{$1$};
 \draw (-6,0) node[below]{$0$};
 \draw[thick] (-6,3)--(10,3);
 \draw[thick, ->] (-6, 0) -- (-6,8) node[left]{$J$};
 \draw (-6,3) node[left]{$J_c$};
 \draw[dashed] plot [smooth] coordinates {(-6, 0.)(-5, 0.252247)(-4, 0.524695)(-3, 0.820862)(-2,1.14528)(-1, 1.5039)(0, 1.90481)(1, 2.35932)(2, 2.88401)(3,3.5046)(4, 4.26413)(5, 5.24333)(6, 6.62345)(7, 8.98277)};
 \draw (6, 6.62) node[left]{$p_B=1-e^{-J}$};
\draw[thick,->] (2.4,3)--(3,3);
\draw[thick,->] (2,3)--(1.4,3);
\draw[thick,->] (2.7, 3.3)--(3,3.5046);
\draw[thick,->] (1.4, 2.6)--(1,2.35932);
 \draw (1.6,3) node[above]{$y^{FK}_p$};
  \draw (4.2,3.5) node[left]{$y^{FK}_t$};
  \draw[fill] (2.2,3) circle(3pt);
 \draw (2.2,2.6) node[below]{FK f.p.};
 \draw[fill] (7,3) circle(3pt);
 \draw (7,2) node[below]{S f.p.};
 \draw[thick,->] (7.8,3)--(7.4,3);
\draw[thick,->] (6.2,3)--(6.6,3);
\draw[dashed] (7,1.8)--(7,4.2);
\draw[thick,->] (7,3.2)--(7,3.6);
\draw[thick,->] (7,2.8)--(7,2.2);
\draw (6.5,3) node[above]{$y^{S}_p$};
  \draw (8,3.5) node[above]{$y^{S}_t$};
\end{tikzpicture}
\caption{RG flows of Potts models in the parameter space $(J, p_B)$ of the Hamiltonian~(\ref{hamDP}), see \cite{Deng2004}.}
\label{fig:phdiagr}
\end{figure}
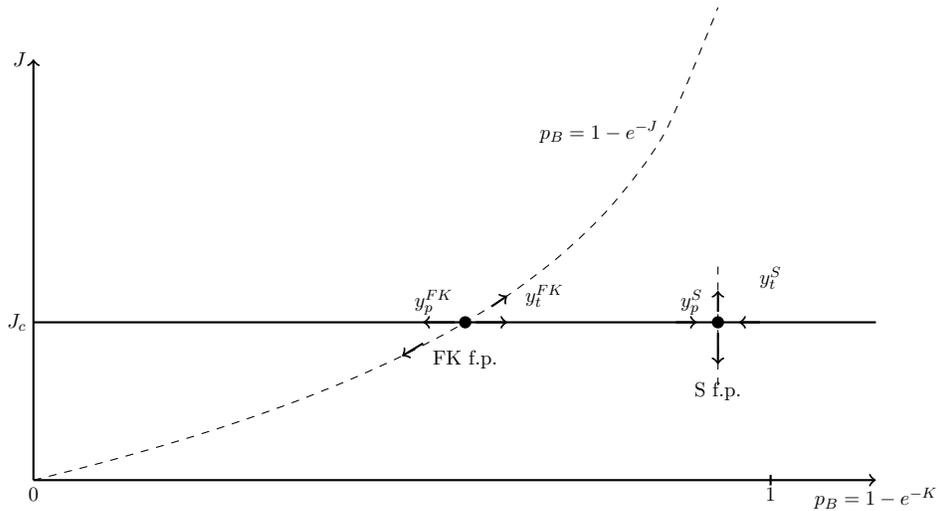

A RG analysis of the Hamiltonian in Eq.~(\ref{hamDP}) of the flow of the  $J$ and $K$ couplings can be performed \cite{CP}, followed by the limit $R\to 1$. One finds two non-trivial fixed points at $J=J_c$,  a repulsive one at $(J=K=J_c)$, describing the FK clusters percolation, and an attractive one $(J=J_c, K=K^*>J_c)$ describing the percolation transitions occurring for $p_B> 1-e^{-J_c}$, see Fig (\ref{fig:phdiagr}). The existence of two different points is clearly shown in our simulations for the $3$-Potts model, see  Fig.~(\ref{FigACQ3}).
The critical point $(J=J_c, K=K^*>J_c)$ attracts therefore the S point. According to this, in the following we can refer to the $S$ clusters as the $p_B$ clusters with 
\begin{align}
K= K^{*},\; p^{\text{S}}_B = 1-e^{-K^{*}}\;:\quad \text{S clusters.}
\label{SC2}
\end{align}
This is for instance what we do for the S clusters in the $Q=3$ Potts model, see Eq.~(\ref{SCQ3}).

\subsection{Partition functions and central charge}
As easily seen from the Eq.~(\ref{hamDP}), the free energy $\mathcal{F}_{R=1}$ does not depend on $K$ and 
coincides with the free energy of the pure $Q-$Potts \cite{Vanderzande92}. Consequently, at $J=J_c$, the universal part of $\mathcal{F}_{R=1}$ on a torus is  $- \log Z_{Q}$, and coincide with the partition function of the S and of the FK points:
\begin{equation}
\label{ZQ}
Z^{\text{S}}=Z^{\text{FK}}= Z_{Q}.
\end{equation}
The above identity implies that the S and FK critical points have the same central charge, which corresponds to the central charge given in the Eq.~(\ref{cq}) for the critical $Q$-Potts . 

\subsection{FK and S clusters fractal dimensions}
The most natural observable for studying a cluster of a given type $X$ is the two-point connectivity:
\begin{equation}
\label{def:2conn}
p^{\text{X}}_{12}({\bf x_1}-{\bf x_2})= \text{Prob.}\left[ \text{${\bf x_1}$ and ${\bf x_2}$ belong to the same X cluster}\right],
\end{equation}
where the vector ${\bf x}$ indicates a point in the bi-dimensional lattice. The $p^{\text{X}}_{12}$ is always considered to be measured at the percolation transition. One of the basic assumptions we make is that the universal behavior of $p^{\text{X}}_{12}$ is captured  by the two-point function of the connectivity field $V^{\text{X}}_{\text{conn}}$:
\begin{equation}
\label{assconn}
p^{\text{X}}_{12}({\bf x_1}-{\bf x_2})=a^{\text{X}}_0 \left<V^{\text{X}}_{\text{conn}}({\bf x_1})V^{\text{X}}_{\text{conn}}({\bf x_2})\right>,
\end{equation}
where $a^{\text{X}}_0$ is a non-universal factor. The above identity has to be understood in the scaling limit and $\left<\cdots\right>$ indicates a correlation function taken in the CFT describing the cluster of type $X$. In the plane (infinite torus) limit, one has: 
\begin{equation}
p^{\text{X}}_{12}({\bf x_1}-{\bf x_2})=\frac{a^{X}_{0}}{| {\bf x_1}-{\bf x_2}|^{4\Delta^{\text{X}}_{\text{conn}}}}, \;\quad \;\text{for} \;|{\bf x_1}-{\bf x_2}|>>1
\end{equation}
where $|{\bf x_1}-{\bf x_2}|$ is the distance between the two points, and  $\Delta^{\text{X}}_{\text{conn}}$ is the conformal dimension of  $V^{\text{X}}_{\text{conn}}$. The $\Delta^{\text{X}}_{\text{conn}}$ yields the fractal dimension $D_f^{\text{X}}$ of the X cluster,  $D_f^{\text{X}}=2-2\Delta^{\text{X}}_{\text{conn}}$. 

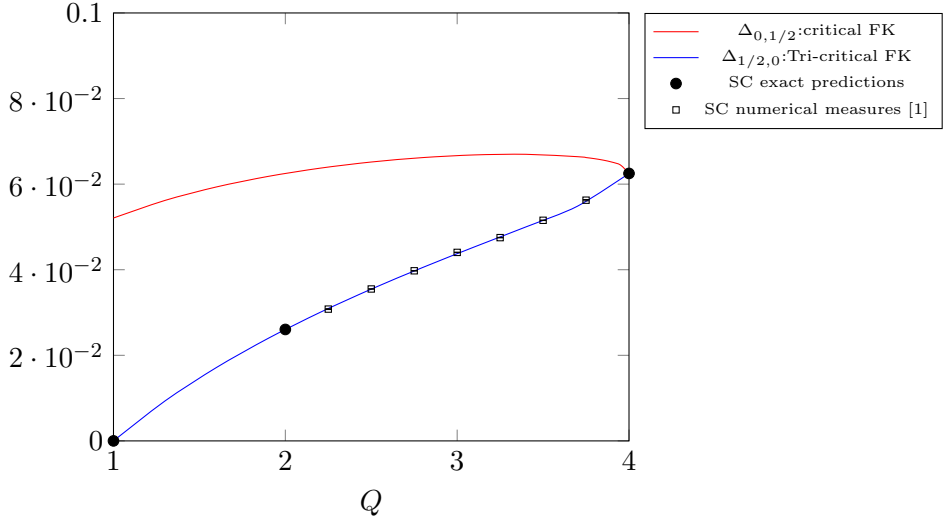
\begin{figure}
\begin{center}
\begin{tikzpicture}[scale=1]
\begin{scope}
\begin{axis}[
title={}, 
	xlabel={$Q$},
	 xtick={1.0,2.0,3.0,4.0},
	xmin=1,
  xmax=4,
  ymin=0,
  ymax=0.1,
   legend pos=outer north east
	]
\addplot[red, smooth,mark= none] 
coordinates{
(1., 0.0520833)(1.3, 0.0562)(1.6, 0.0593526)(1.9, 0.0618081)(2.2, 0.0637179)(2.5, 0.0651691)(2.8, 0.0662045)(3.1,0.0668257)(3.4, 0.0669717)(3.7, 0.0664153)(3.75, 0.0662079)(3.8, 0.0659456)(3.85, 0.0656078)(3.9, 0.0651544)(3.95, 0.0644816)(4,0.0625)
};

\addplot[blue, smooth,mark= none] 
coordinates{
(1., 0)(1.3, 0.00934237)(1.6, 0.0171716)(1.9,0.0239582)(2.2, 0.0299904)(2.5, 0.0354652)(2.8, 0.0405324)(3.1, 0.045326)(3.4, 0.050001)(3.7, 0.0548431)(4,0.0625)
};

\addplot[only marks, mark=*, fill=black] 
coordinates{
(1., 0)(2, 0.0260417)(4,0.0625)
};
\addplot[only marks, mark=square,  mark size=1.2pt, fill=blue,
error bars/.cd, y dir=both, y explicit ] 
coordinates{
(2.25, 0.0308) +- (0.0, 0.0001) (2.5, 0.0355) +- (0.0,0.00005) (2.75, 0.03975) +- (0.0,0.00005)  (3.0, 0.04405) +- (0.0,0.00005) (3.25, 0.0475) +- (0.0,0.00005) (3.5, 0.05155) +- (0.0,0.00005) (3.75,0.05625) +- (0.0, 0.00015)
};

\legend{{\tiny $\Delta_{0,1/2}$:critical FK}, {\tiny $\Delta_{1/2,0}$:Tri-critical FK}, {\tiny SC exact predictions},{\tiny SC numerical measures \cite{Delfino13}} }
\end{axis}
\end{scope}
\end{tikzpicture}
\caption{Connectivity fields dimensions as a function of $Q \in[1,4]$.}
\label{FKSCdf}
\end{center}
\end{figure}

Via the Coulomb gas approach \cite{Nienhuis82}, the dimension of $V^{\text{FK}}_{\text{conn}}$ and of $V^{\text{tri-FK}}_{\text{conn}}$ have been determined:
\begin{equation}
\label{connfk}
\Delta_{\text{conn}}^{\text{FK}}= \Delta_{0,\frac12}, \quad \Delta_{\text{conn}}^{\text{tri-FK}}= \Delta_{\frac12,0}. 
\end{equation}
The values of the above dimensions as a function of $Q \in[1,4]$, are plotted in Fig.~(\ref{FKSCdf}). Note that $\Delta_{\text{conn}}^{\text{tri-FK}}$ corresponds to the continuation $\Delta_{\text{conn}}^{\text{FK}}$ on the tri-critical branch. 

The  $\Delta^{S}_{\text{conn}}$ of Ising S cluster has been determined in \cite{SV}. For $Q = 2$, where the spins $s(x)$ play the role 
of vacancies degree of freedom, the model in Eq.~(\ref{hamDP}) becomes a dilute $R-$state Potts model 
of the type defined in Eq.~(\ref{def:dilPotts}) \cite{CP}. The Ising S point is described by the tri-critical $(R=)1-$Potts point. Consistently with Eq.~(\ref{ZQ}), one can verify that $Z_{Q=2}=Z_{Q=1}^{\text{tri}}$. The Ising S connectivity field corresponds therefore to the one of the tri-critical $1$-Potts, i.e.  
$\Delta_{\text{conn}}^{\text{S}}=\Delta_{\text{conn}}^{\text{tri-FK}}=\Delta_{1/2,0}=187/96$ at $Q=2$ ($\beta^2=3/4$).

A neat relation between the S point and the tri-critical $Q$-Potts is not available for $Q\neq 2$. For $Q=1$, where one S cluster covers 
the whole lattice, one expects  $\Delta_{\text{conn}}^{\text{S}}=0$, i.e. $D_f^{\text{S}}=0$. For $Q=4$, the marginality of the FK pivotal exponent 
$y^{\text{FK}}_{p_B}$ , see Eq.~(\ref{FKexp}), hints to the coalescence of the FK and S fixed points \cite{Vanderzande92}. 
This implies $\Delta_{\text{conn}}^{\text{S}}=1/16$ at $Q=4$. 
On the basis of these two observations, the conjecture: 
\begin{equation}
\label{spinconn}
\Delta_{\text{conn}}^{\text{S}}=\Delta_{\text{conn}}^{\text{tri-FK}}=\Delta_{\frac12,0},
\end{equation}
according to which the S clusters have the same fractal dimension as the tri-critical FK clusters, was made\cite{Vanderzande92}. A direct numerical 
verification of the validity of  Eq.(\ref{spinconn}) was done for $Q=3$ in \cite{Vanderzande92} and for non-integer $Q$ in \cite{Delfino13}, 
see Fig.~(\ref{FKSCdf}). This conjecture was for instance assumed true in \cite{Deng2004,Janke2004}. 

The three-point connectivity $p_{123}^{\text{X}}$ is defined as 
\begin{equation}
\label{def:3conn}
p^{\text{X}}_{123}({\bf x_1},{\bf x_2},{\bf x_3})= \text{Prob.}\left[\text{${\bf x_1}$,${\bf x_2}$,${\bf x_3}$ belong to the same X cluster}\right].
\end{equation}
In the scaling limit, by using the $SL(2,\mathbb{C})$ (global conformal) invariance, one obtains: 
\begin{equation}
\label{def:3conn_2}
p^{\text{X}}_{123}(0,1,\infty)= \left<V_{\text{conn}}^{\text{X}}(0)V_{\text{conn}}^{\text{X}}(1)V_{\text{conn}}^{\text{X}}(\infty)\right>=C_{\Delta_{\text{conn}}^{\text{X}},\Delta_{\text{conn}}^{\text{X}}}^{\Delta_{\text{conn}}^{\text{X}}}.
\end{equation}
Note that in the above equation we are considering the limit of the infinite plane. The $p_{123}^{\text{tri-FK}}$ is expected to be given by the Liouville constants defined in Eq.~(\ref{Liouc}):
\begin{equation}
\label{3-triFKLi}
C_{\Delta_{\text{conn}}^{\text{tri-FK}},\Delta_{\text{conn}}^{\text{tri-FK}}}^{\Delta_{\text{conn}}^{\text{tri-FK}}}=\left(C^{c\leq 1}\right)_{(1/2,0)(1/2,0)}^{(1/2,0)} \; .
\end{equation}
\noindent In \cite{Delfino13} it was numerically observed that: 
\begin{equation}
\label{3-S}
C_{\Delta_{\text{conn}}^{\text{S}},\Delta_{\text{conn}}^{\text{S}}}^{\Delta_{\text{conn}}^{\text{S}}}\begin{cases} &= \left(C^{c\leq 1}\right)_{(1/2,0)(1/2,0)}^{(1/2,0)}, \quad \text{for}\; Q= 2 \\
&\neq \left(C^{c\leq 1}\right)_{(1/2,0)(1/2,0)}^{(1/2,0)}\quad \text{for}\; Q \neq  2
\end{cases}
\; .
\end{equation} 
The S and the tri-critical FK thus do not belong to the same universality class. In particular the connectivity fields have the same conformal dimension but are different fields for $Q\neq 2$:
\begin{equation}
\Delta_{\text{conn}}^{\text{S}}=\Delta_{\text{conn}}^{\text{tri-FK}},\quad \begin{cases} V_{\text{conn}}^{\text{S}}=V_{\text{conn}}^{\text{tri-FK}}& \text{for}\; Q=2 \\
V_{\text{conn}}^{\text{S}}\neq V_{\text{conn}}^{\text{tri-FK}}& \text{for}\; Q\neq2 
\end{cases}
\; .
\end{equation}

\subsection{Thermal and pivotal exponents} 
We denote by $y^{\text{X}}_J$ and $y^{\text{X}}_{p_B}$ the exponents associated to the $X=FK,S$ fixed point, that determine the RG flow in the  ($J$,$p_B$) plane. The FK point is fully repulsive and therefore $y^{FK}_J$ and  $y^{FK}_{p_B}$ are positive. The S point is repulsive in the $J$ direction and attractive in the $p_B$ one, such that $y^{\text{S}}_{J}>0$ and $y^{\text{S}}_{p_B}<0$.  

The $y^{\text{FK}}_{J}$ is related to the FK-CFT energy field $\varepsilon^{\text{FK}}$ of dimension $\Delta_{\varepsilon}^{\text{FK}}$ and $y^{FK}_{p_B}$ coincides with the fractal dimension $D^{\text{piv}}_{f}= 2-2\Delta_{\text{piv}}^{\text{FK}}$ of the pivotal bonds (or red bonds)\cite{Coniglio89}: 
\begin{equation}
\label{FKy}
y^{\text{FK}}_{J}= 2 - 2\Delta_{\varepsilon}^{\text{FK}}, \quad y^{\text{FK}}_{p_B}=2-2\Delta_{\text{piv}}^{\text{FK}} \; .
\end{equation}
The conformal dimension $\Delta_{\varepsilon}^{\text{FK}}$ and $\Delta_{\text{piv}}^{\text{FK}}$  have been computed via Coulomb gas methods: 
\begin{equation}
\label{FKexp}
 \Delta_{\varepsilon}^{\text{FK}}=\Delta_{1,2}, \; \Delta_{\text{piv}}^{\text{FK}}=\Delta_{2,0}\; .
\end{equation}

Concerning the exponent $y^{\text{S}}_{p_B}$, according to the S/tri-critical $1$-Potts identification, one expects, for $Q=2$, $\Delta_{\text{piv}}^{\text{S}}= \Delta_{\text{piv}}^{\text{tri-FK}}=\Delta_{0,2}=21/16$. 
For $Q\neq 2$ the following identification has been suggested in \cite{Deng2004}~:
\begin{equation}
\label{ypb}
\Delta_{\text{piv}}^{\text{S}}= \Delta_{\text{piv}}^{\text{tri-FK}}\to  y^{\text{S}}_{p_B}= 2- 2\Delta_{0,2}\; .
\end{equation}
By studying the wrapping probability, we have verified that the above identification holds for $Q=3$.

In the following section we will discuss the thermal exponents $y^{\text{S}}_{J}$ and the associated energy field $\varepsilon^{\text{S}}$, pointing out the main results concerning this field.

\section{The thermal exponent and the energy field in the S fixed point: new results}

We focus here on the exponent $y^{\text{S}}_{J}$ associated to the S point, in particular for $Q\neq 2$. We denote $\varepsilon^{\text{S}}$ the energy field, of conformal dimension $\Delta^{\text{S}}_{\varepsilon}$, which is associated to $y^{\text{S}}_{J}$
\begin{equation}
y^{\text{S}}_{J}= 2- 2\Delta^{\text{S}}_{\varepsilon}.
\end{equation}
For $Q=2$, $y^{\text{S}}_{J}$ has been computed in \cite{SV} using the S/tri-critical $1$-Potts identification. In the tri-critical $1$-Potts model the thermal sector contain two relevant fields, of 
dimensions $\Delta_{2,1}=1/16$ and $\Delta_{3,1}=\Delta_{1,2}=1/2$, that are respectively associated to the $\mu$ and to the $J$ fields in Eq.~(\ref{def:dilPotts})\cite{SV}. The energy field 
$\varepsilon^{\text{S}}$ has therefore dimension $\Delta_{3,1}$ and $y^{\text{S}}_{J}=y^{\text{FK}}_{J}$ at the $Q=2$ point. This result has been first  predicted in \cite{Coniglio80} by an RG calculation in which the flow of $J$ is decoupled from $p_B$. 

\begin{figure}
\begin{center}
\begin{tikzpicture}[scale=1]
\begin{scope}
\begin{axis}[
title={Energy fields dimension},
	xlabel={$Q$},
	 xtick={1.0,2.0,3.0,4.0},
	xmin=1,
  xmax=4,
  legend pos=outer north east
	]
\addplot[red, smooth,mark= none] 
coordinates{
(1.,0.625)(1.3,0.582109)(1.6,0.544637)(1.9,0.510729)(2,0.5)(2.2,0.479186)(2.5,0.449108)(2.8,0.419687)(3,0.4)(3.1, 0.390011)(3.4,0.358705)(3.7,0.32264)
(3.8, 0.308001)(3.9, 0.289924)(3.95, 0.277736)(4,0.25)};
\addplot[black, smooth,mark= none] 
coordinates{
(1.,0)(1.3, 0.0198181)(1.6, 0.0384647)(1.9,0.0565292)(2.2, 0.0744566)(2.5, 0.0926615)(2.8, 0.111621)(3.1, 0.132015)(3.4, 0.155056)(3.7, 0.183774)(3.75, 0.189677)(3.8, 0.196163)(3.85, 0.203476)(3.9, 0.212094)(3.95, 0.223253)(4,0.25)};
\addplot[blue, smooth,mark= none] 
coordinates{
(1., 0.333333)(1.3, 0.386182)(1.6, 0.435906)(1.9, 0.484078)(2,0.5)(2.2, 0.531884)(2.5, 0.580431)(2.8, 0.63099)(3.1, 0.685372)(3.4, 0.746817)(3.7, 0.823398)(3.8, 0.856434)(3.85, 0.875935)(3.9, 0.898917)(3.95, 0.928675)(4,1)};
%
\addplot[only marks, mark=*, mark size=1pt, 
error bars/.cd, y dir=both, y explicit ] 
coordinates{
(1.25, 0.635) +- (0.0, 0.025) (1.5, 0.595) +-  (0.0,0.01) (2, 0.5095) (0.0,0.0015) (2.5, 0.4552) (0.0, 0.0022) (3, 0.4045) (0.0,0.0005) 
};
\addplot[green!100,only marks, mark=square] 
coordinates{
(3, 0.4)
};
%
\legend{{\tiny $\Delta_{1,2}$: FK energy}, {\tiny $\Delta_{2,1}$:Tri-crit. FK energy 1}, {\tiny $\Delta_{3,1}$:Tri-crit. FK energy 2},{\tiny $\varepsilon^{\text{S}}$ from torus $p^{S}_{12}$},{\tiny $\varepsilon^{\text{S}}$ from wrapping}}
\end{axis}
\end{scope}
\end{tikzpicture}
\caption{Plot of relevant thermal dimensions for the critical and tri-critical Potts model as a function of $Q$. Dots are numerical measurments of $\Delta_\varepsilon^{\text{S}}$ from Tab.~\ref{Tablea}  and the green square is the value obtained in the Eq.~(\ref{Finaly}).}
\label{endf}
\end{center}
\end{figure}
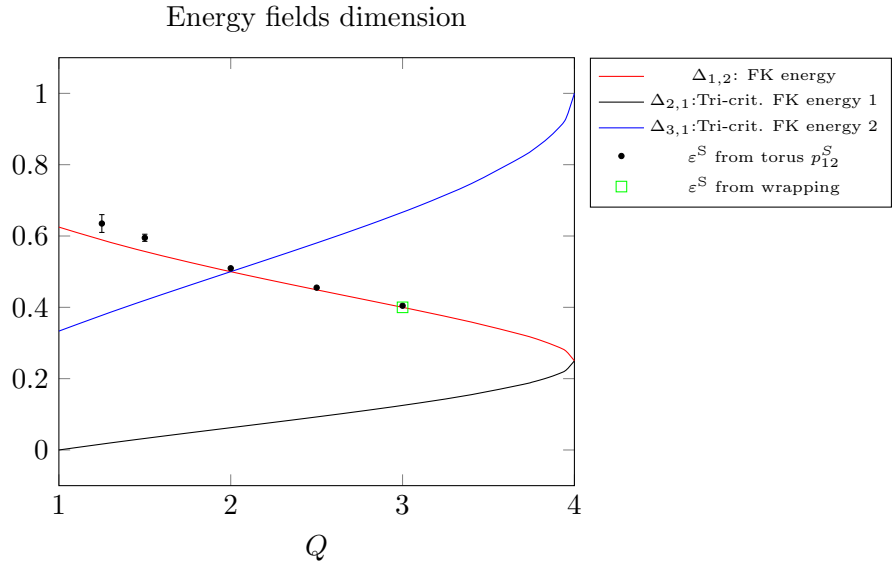

Can we extend the $Q=2$ result to the other values of $Q$? 
In Fig.~(\ref{endf}) we plotted for $Q\in [1,4]$ the values of  $\Delta_{1,2}$, $\Delta_{2,1}$ and $\Delta_{3,1}$, that are respectively the conformal dimension of $\varepsilon^{\text{FK}}$ 
and of the leading and the sub-leading energy fields in the tri-critical Potts. In the same Fig.~(\ref{endf}), we show the values for $\Delta_{\varepsilon}^{\text{S}}$ that we evaluated numerically 
by studying the finite effects of two different observables, the $p^{\text{S}}_{12}$ connectivity and the wrapping probability. 

On the basis of numerical results, see Table \ref{Tablea} and Eq.~(\ref{Finaly}), we can conclude that:
\begin{equation}
\label{yj}
\Delta_{\varepsilon}^{\text{S}}= \Delta_{\varepsilon}^\text{FK}, \quad \text{for}\; Q\in [1,4] \; .
\end{equation}
So, differently from $\Delta^{\text{S}}_{\text{conn}}$ and $\Delta^{\text{S}}_{\text{piv}}$, see Eq.~(\ref{spinconn}) and Eq.~(\ref{ypb}), the $\Delta_{\varepsilon}^{\text{S}}$ follows the critical FK value $\Delta_{\varepsilon}^{\text{FK}}$. 

According to the Monte Carlo measures of $p_{12}^{\text{S}}$, see Table~\ref{TableQ2} and Table~\ref{TableQ3}, we conjecture that:
\begin{equation}
\label{entor}
\left< \varepsilon^{\text{S}} \right> = \left< \varepsilon^{\text{FK}} \right>, \quad \text{for}\; Q\in [1,4]
\end{equation}
where $\left< \varepsilon^{\text{X}} \right>$ is the torus one-point function of the $\varepsilon^{\text{X}}$ field.

Despite the fact that the fields $\varepsilon^{\text{S}}$ and $\varepsilon^{\text{FK}}$ have the same dimension and the same one-point torus function, they are not the same field, as inferred from our numerical findings.  On the one hand, the representation associated to $\varepsilon^{\text{FK}}$ has a vanishing singular state at level two. The corresponding fusion rules, see the Eq.~(\ref{fus2lev}) implies:
\begin{equation}
\label{enfkope}
C_{\Delta,\Delta}^{\Delta^{\text{FK}}_{\varepsilon}}=\left< V_{\Delta}V_{\Delta}\;\varepsilon^{\text{FK}}\right>= 0 \quad \text{if} \; \Delta \neq \Delta_{0,1/2}\; .
\end{equation}
On the other hand, see the Sec.~\ref{coefc} and in particular the Eq.~(\ref{CCC}), the comparison between our CFT predictions and the numerical data yields:  
\begin{equation}
\label{3ens}
C_{\Delta^{\text{S}}_{\text{conn}},\Delta^{\text{S}}_{\text{conn}}}^{\Delta^{\text{S}}_{\varepsilon}}=\left<V^{\text{S}}_{\text{conn}}V^{\text{S}}_{\text{conn}}\;\varepsilon^{\text{S}}\right> \neq 0 \quad \text{for} \; Q\in [1,4].
\end{equation}
For $Q=2$, the following identity:
\begin{equation}
\label{3ensq2}
C_{\Delta^{\text{S}}_{\text{conn}},\Delta^{\text{S}}_{\text{conn}}}^{\Delta^{\text{S}}_{\varepsilon}}=\left(C^{c\leq 1}\right)_{(1/2,0)(1/2,0)}^{(3,1)}\quad \text{for} \; Q=2, 
\end{equation}
where $C^{c \leq 1}$ is given in Eq.~(\ref{Liouc}), is in good agreement with the numerical findings. 
In summary, on the basis of our findings, we conclude that:
\begin{equation}
\Delta^{\text{S}}_{\varepsilon}=\Delta_{\varepsilon}^{\text{FK}},\quad \begin{cases} \varepsilon^{\text{S}}=\varepsilon^{\text{tri-FK}}& \text{for}\; Q=2 \\
\varepsilon^{\text{S}}\neq \varepsilon^{\text{FK}}\neq \varepsilon^{\text{tri-FK}}& \text{for}\; Q\in [1,4], Q\neq 2
\end{cases}
\; .
\end{equation}
\section{Finite size corrections to two-point connectivity: CFT predictions}
\label{sec:2pc}
In numerical simulations it is often convenient to consider lattices with periodic boundary conditions on both horizontal and vertical directions. An $L_{h}\times L_{v}$ square lattice with these boundary conditions has the same topology as a torus of nome
\begin{equation}
\label{nome}
q= e^{-2 \pi \frac{L_{h}}{L_{v}}}.
\end{equation}
When one measures the two-point connectivity of any type of clusters, see Eq.~(\ref{def:2conn}), the topological universal effects 
become manifest when the distance between the points starts to be of the order of the lattice size.  
In some recent works \cite{Javerzat19,Javerzat19b} we have focused on the FK cluster and shown that these topological effects are   excellent observables to evaluate the FK-CFT structure constants. Here we apply the same analysis to the S clusters.

We make three main assumptions. The first two assumptions are more general and concern the fact that the $p_{12}^{\text{X}}$ can be studied by correlations of local fields in a CFT. 
More specifically, we assume that, in the scaling limit, i) the S clusters are conformally invariant and ii)  the $p_{12}^{\text{X}}$ is described by a two point conformal correlator, see Eq.~(\ref{assconn}). On the basis of these assumptions, the topological effects can be expressed in terms of the expansion\cite{Javerzat19, Javerzat19b}:
\begin{align}
\label{eq:genss}
p^{\text{X}}_{12}({\bf x})&=\frac{a_0^{\text{X}}}{r^{4\Delta^{\text{X}}_{\text{conn}}}} \; \sum_{\substack{(\Delta, \bar{\Delta})\\\ \Delta\geq \bar{\Delta}}}\; (2-\delta_{\Delta,\bar{\Delta}})\; \left<V_{\text{conn}}^{\text{X}}V_{\text{conn}}^{\text{X}} V_{(\Delta, \bar{\Delta})} \right>\left<V_{(\Delta, \bar{\Delta})}\right>\times\nn \\
&\times \cos\left((\Delta-\bar{\Delta})\;\theta\right) \left(\frac{r}{L_v}\right)^{\Delta+\bar{\Delta}},
\end{align} 
where 
\begin{equation}
{\bf x}= (r \cos \theta, r \sin \theta)= r\; e^{i \theta},
\end{equation}
and $\left<\cdots\right>$ is the torus one-point function in the X-CFT, X=FK or S.
The sum runs over the  fields $V_{(\Delta, \bar{\Delta})}$ of left and right dimension $(\Delta, \bar{\Delta})$. In the above expression, 
the field $V_{(\Delta, \bar{\Delta})}$ can indicate a descendant, like for instance the stress energy tensor components $T$ and $\bar{T}$, 
which have dimension $(\Delta, \bar{\Delta})=(2, 0)$ and  $(\Delta, \bar{\Delta})=(0, 2)$. Even though it is not manifest in the above 
expression, the connectivity  is (modular) invariant under the replacement $L_{v}\to L_{h}$.

The third assumption we make is that the identity field ($\Delta=\bar{\Delta}=0$), the energy field $\varepsilon^{\text{X}}$, and the stress-energy tensor fields $T^{\text{X}}$, $\bar{T}^{\text{X}}$ are the dominant fields, i.e. with the lowest conformal dimension, contributing to the expansion Eq.~(\ref{eq:genss}).
This assumption was tested valid for the FK cluster in \cite{Javerzat19}. Using this third assumption, the Eq.~(\ref{eq:genss}) takes the form:
\begin{equation}\label{predperco}
p^{\text{X}}_{12}({\bf x}) = \frac{a^{\text{X}}_0}{r^{4 \Delta^{\text{X}}_{\text{conn}}}}\left(1+c^{\text{X}}_{\varepsilon}\left(q\right)\left(\frac{r}{L_{h}}\right)^{2\Delta^{\text{X}}_{\varepsilon}}+2c^{\text{X}}_{T}\left( q \right)\cos(2\theta)\left(\frac{r}{L}\right)^{2}+o\left( \left(\frac{r}{L_h}\right)^2\right)\right),
\end{equation}
where:
\begin{align}
\label{CFTcoeff}
c^{\text{X}}_{\varepsilon}(q) &=(2\pi)^{2\Delta^{X}_{\varepsilon}}\; C_{\Delta^{\text{X}}_{\text{conn}},\Delta^{\text{X}}_{\text{conn}}}^{\Delta^{\text{X}}_{\varepsilon}} \;\left<\varepsilon^{\text{X}}\right>,\; c^{\text{X}}_{T}(q) = \frac{2\Delta^{\text{X}}_{\text{conn}}}{c} \left< T^{\text{X}}\right>,\;\left< T^{\text{X}}\right>=-2\pi^2 \partial_q \log{ Z^{\text{X}}(q)}.
\end{align} 
The dependence on the torus shape, i.e. on the nome $q$ in Eq.~(\ref{nome}), is contained in the torus one point function $ \left<\varepsilon^{\text{X}}\right>$ and $ \left< T^{\text{X}}\right>$.    
\section{Measurements of S cluster two-point connectivity}
\label{Meas}

Our aim is to compare the predictions in Eq.~(\ref{predperco}) with the Monte Carlo measurements. We determine
\begin{itemize}
\item the dominant topological correction for $p_{12}^{S}$ and show that it is given by $(r/L_h)^{2 \Delta_{1,2}}$. This result supports  the Eq.~(\ref{yj}) and the Eq.~(\ref{3ens}).
\item  the coefficient $c^{\text{S}}_\varepsilon(q)$. This will allow us to test the Eq.~(\ref{entor}) and the Eq.~(\ref{3ensq2}).
\item determine the form of the coefficient $c_T(q)$. This will further confirm the validity of Eq.~(\ref{predperco}) and therefore of our assumptions. 
\end{itemize}

We performed numerical simulations for the Potts model on a $L_h \times L_v$ the square lattice periodic with toric boundary conditions. We simulated systems of size 
$L_h=L_v = 8192$ or $L_v = 2048, L_h=\alpha\; L_v$ with $\alpha \geq 1$ the aspect ratio. We employed the Wolff algorithm \cite{Wolff} for integer values of $Q$ and 
the Chayes-Machta algorithm \cite{Chayes98} for non integer values of $Q$, which is a generalisation of the Swendsen-Wang algorithm, 
see~\cite{Deng07} for details. For each value of $Q$ and sizes, we simulated one million independent samples. 

The $p_{12}^{\text{X}}({\bf x})$ are measured separately for:
\begin{align}
{\bf x} &= (r,0)=r,\quad \text{(the two points are aligned along the horizontal axis)} \nonumber \\
{\bf x} &= (0,r)= i r,\quad \text{(the two points are aligned along the vertical axis)}.
 \end{align}
In order to isolate the different contributions, it is convenient to analyse the sum and the difference between these two quantities. The CFT predictions in Eq.~(\ref{predperco}) yield:
\begin{align}
\label{sum2p}
r^{4 \Delta_{\text{conn}}^{\text{X}}} \left(p^{\text{X}}_{12}(r) + p^{\text{X}}_{12}(i r)\right) &= a^{\text{X}}_0\left( 2 + 2 \;c_{\varepsilon}^\text{X}\left(q\right)\left(\frac{r}{L_{h}}\right)^{2\Delta_{1,2}} 
+o\left( \left(\frac{r}{L_h}\right)^2\right)\right) \\
\label{diff2p}
r^{4 \Delta_{\text{conn}}^{\text{X}}} \left(p^{\text{X}}_{12}(r) - p^{\text{X}}_{12}(i r)\right) &= a^{\text{X}}_0\left( 4 \;c_{T}^\text{X}\left( q \right) \left(\frac{r}{L_{h}}\right)^{2}+o\left( \left(\frac{r}{L_h}\right)^2\right)\right)  \; .
\end{align}

\subsection{The dominant correction: the energy field}
\begin{figure}[!ht]
\begin{center}
\includegraphics[width=7.5cm]{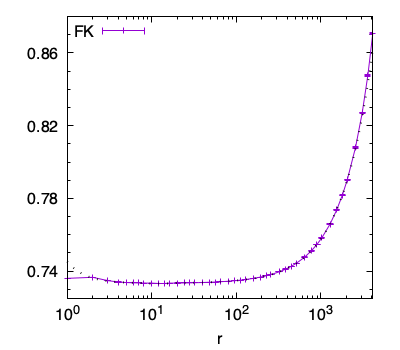}
\includegraphics[width=7.5cm]{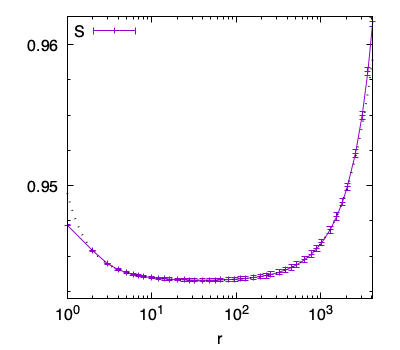}
\caption{\small 
$r^{4 \Delta_{\text{conn}}^{\text{X}}} p^{\text{X}}_{12}(r)$  vs $r$ for $Q=1.25$, FK clusters on the left and S clusters on the right.
}
\label{Q125}
\end{center}
\end{figure}

In Fig.~\ref{Q125}, we show measured values of $r^{4 \Delta_{\text{conn}}^{\text{FK}}} p^{\text{FK}}_{12}(r)$ and of $r^{4 \Delta_{\text{conn}}^{\text{S}}} p^{\text{SC}}_{12}(r)$ as a function of 
$x$ for $Q=1.25$  and $L=L_h=L_v=8192$. In this case, i.e. the one of a square domain, the connectivity depends only on the distance between the two marked points and not on the orientation of the line segment linking them with respect to the domain axes.%
There are two important points to notice. First, error bars are much larger for the S clusters. This is due to the implementation of the Chayes-Machta algorithm. While the measurement is done on all the clusters of the lattice for the FK clusters, they are done only on a fraction $1/1.25$ for the S clusters. Moreover, this fraction is not fixed, it is only an average fraction. Over the samples, this quantity fluctuates, which produces additional errors on the final measurement. 
The second point is that the correction to the scaling (corresponding to the deviation from a plateau) is much smaller for the S clusters compared to the FK clusters. For $Q=1.25$, they are ten times smaller. This explains why the measurements for S clusters will produce less precise results. 

The Fig.~\ref{Q125} contains also a best fit to the form 
\beq
r^{4 \Delta_{\text{conn}}^{\text{FK}}} p^{\text{FK}}_{12}(r) = \left(1 + \frac{e_1^{\text{FK}}}{r} + \frac{e_2^{\text{FK}}}{r^2} \right) a_0^{\text{FK}} \left( 1 + c^{\text{FK}} \left(\frac{r}{L}\right)^{2 \Delta^{\text{FK}}}\right) \;
\eeq
for the FK correlation function and 
\beq
r^{4 \Delta_{\text{conn}}^{\text{S}}} p^{\text{S}}_{12}(r) = \left(1 + \frac{e_1^{\text{S}}}{r} + \frac{e_2^{\text{S}}}{r^2} \right) a_0^{\text{S}} \left( 1 + c^{\text{S}} \left(\frac{r}{L}\right)^{2\Delta^{\text{S}}}\right) \;
\eeq
for S correlation function. $e_1$ and $e_2$ are free parameters to take into account the small distances corrections. For more details on these fits, see \cite{Picco16,Picco19}.
The fits are done for $r \in [6:L/4]$.  
The result of the best fit for the $FK$ correlation function is 
\beq
a_0^{\text{FK}} = 0.73323 (1) \; ; \; c^{\text{FK}} = 0.397 (1) \; ; \;  2\Delta^{\text{FK}}  = 1.177 (1) \; ,
\eeq 
with the error in parenthesis. The agreement for  $2\Delta^{\text{FK}}$ is good with $2\Delta_{1,2} = 1.1776$.
A similar best fit for the $SC$ correlation function gives
\beq
a_0^{\text{S}} = 0.94320 (1) \; ; \; c^{\text{S}} = 0.042 (2) \; ; \;  2\Delta^{\text{S}}  = 1.27 (5) \; .
\eeq 
Here again the measured value $2\Delta^{\text{S}}$ is close to $2\Delta_{1,2}$. The agreement is less good than for the FK correlation function 
but note that the correction is ten times smaller \ie\ $c^{\text{S}} \simeq c^{\text{FK}}/10$. 

\begin{table}[h]
\centering
\begin{tabular}{ | c || c | c  || c ||  c | c ||   } 
\hline
   $Q$   &   $c^{\text{FK}}$   &  $2\Delta^{\text{FK}}  $ & $2\Delta_{1,2}$ &  $c^{\text{S}}$  &  $2\Delta^{\text{S}}$ \\
\hline
  1.25     &     0.397 &  1.177 (2 )  &  1.1776    &    0.042 (2) &   1.27 (5) \\ 
  1.5     &     0.434  &  1.113 (3)  &   1.1133    &    0.083      &    1.19 (2)    \\
   2      &      0.495  &  1.004 (2 )   &   1.0       &  0.168        &    1.019 (3)  \\
   2.5   &      0.549  &   0.901  &    0.8982       &  0.263      &     0.910 (5)  \\
   3      &      0.604  &   0.805 &    0.8             &  0.375        &     0.809    \\
   \hline
\end{tabular}
\caption{$c^{\text{X}}$ and $\delta^{\text{X}}$ for various $Q$. Error bar is shown in parenthesis and is less or equal to one of the last digit if not shown.}
\label{Tablea}
\end{table}
The same measurements and fits have also been made for $Q=1.5, 2, 2.5$ and $3$. For $Q=2.5$ and $3$, we obtained the S clusters by using 
the value $K^*= 3 J_c$ and $K^*=1.7 J_c$ as determined in \cite{Delfino13}. Note that it corresponds to an attractive fixed point, so a very precise determination for $K^*$ is not necessary. 
For $Q=1.5$ and $Q=2$, we just considered $K^* = \infty$ as for $Q=1.25$.  In Tab.~\ref{Tablea}, 
we reported the values we obtained for $c^{\text{X}}$ and $2\Delta^{\text{X}}$ for X= FK, S. The agreement between $2\Delta^{FK}$ and 
$2\Delta_{1,2}$ (also shown in the table) is always very good. 
It is slightly less good between $2\Delta^{S}$ and $2\Delta_{1,2}$ but again, the corrections are always smaller as seen by the values of $c^{\text{S}}$. 
Last, note that the agreement is less good for $Q=3$ for both the FK clusters and the S clusters. This was already noted for the FK clusters in \cite{Javerzat19} where it was explained by the fact that the next leading contribution by the thermal fields with dimension $\Delta_{1,3}$ becomes non-negligeable for $Q \geq 3$. 

The conclusion is that our measurements are in agreement with the claim that the correction is proportional
to $(r/L)^{2\Delta_{1,2}}$ for the FK and the S clusters.

\subsection{The coefficient ${\bf c^{\text{X}}(q)}$: the energy field correlation functions}
\label{coefc}
We focus now on the coefficient  $c^{\text{X}}(q)$, and in particular in its dependence on the torus shape parameter $\alpha$:
\begin{equation}
\label{alpha}
L_{h} = \alpha \; L_v, \quad q(\alpha)=e^{-2\pi \alpha}
\end{equation}
From the CFT predictions in the Eq.~(\ref{CFTcoeff}), the $c^{\text{X}}(q)$  is expected to be factorized in a $q$-independent part, associated to the structure constant, 
and a $q$-dependent one, associated to $\left<\varepsilon^{\text{X}}\right>$.
This allows us to compare directly the $q$-dependence of the functions $\left<\varepsilon^{\text{FK}}\right>$ and $\left<\varepsilon^{\text{S}}\right>$. In order to split the two contributions,  we have first determined numerically the ratio 
$c^{\text{X}}(q(\alpha))/c^{\text{X}}(q(1))$ for both the type of clusters, X=FK, S.

We have measured these quantities for $Q=2$ and $Q=3$ with $L_h = 2048$ and various values of $\alpha$.
The results of the measurements are shown in Tab.~\ref{TableQ2} for $Q=2$ and in Tab.~\ref{TableQ3} for $Q=3$ as well as the theoretical prediction for $\left<\varepsilon^{\text{FK}}\right>_{q=e^{-2\pi \alpha}}/\left<\varepsilon^{\text{FK}}\right>_{q=e^{-2\pi}}$. The $\left<\varepsilon^{\text{FK}}\right>$ for general value of the central charge was computed in \cite{Zuber89} where an expression in term of a bi-dimensional integral over the torus was given. Here we compute the same function using a different approach, finding a more compact expression, see Appendix \ref{efktorus}. The agreement between the FK and the S ratios  is near perfect, besides the point $Q=3$ and small $\alpha$. Again, this is expected\cite{Javerzat19} due to the additional corrections from the channel $(1,3)$ for $Q \geq 3$. These contributions influence the measurement in particular for small values of $\alpha$. So, we can conclude that our measures strongly supports the Eq.~(\ref{entor}). This result is not completely surprising. Indeed, we recall that the Eq.(\ref{ZQ}) is a consequence of the fact that the free energy $\mathcal{F}_{R=1}$, see the Eq.~(\ref{hamDP}), is the same for the FK and S point. The $\left<\varepsilon^{\text{X}}\right>$, $X=FK,S$, which are given by the derivative of $\mathcal{F}_{R=1}$ with respect to $J$, are then expected to be the same. The difference between the FK and S points is in their percolative behavior, so when, for instance, connectivity fields are involved, as we discuss below.

\begin{table}[h]
\centering
\begin{tabular}{ | c || c | c | c | c | c || c ||   } 
\hline
   $\alpha$       &    1.5             &       2             &       3           &       4        &        6        \\ 
\hline
  $\text{FK}$     &    0.7579 (6) &    0.5529  (7) &  0.2757 (9)  &  0.1394 (6)  &   0.0308 (7)  \\ 
  $\text{S}$     &    0.757  (2)  &    0.550 (2)    &  0.275   (3)  &  0.139   (2)   &  0.028  (2)  \\ 
  $\frac{\left<\varepsilon^{\text{FK}}\right>_{q=e^{-2\pi \alpha}}}{\left<\varepsilon^{\text{FK}}\right>_{q=e^{-2\pi}}}$  &    0.7556       &    0.5526       &  0.2780       &   0.1330   &  0.0286  \\ 
  \hline
\end{tabular}
\caption{$c^{\text{X}}(q(\alpha))/c^{\text{X}}(q(1))$ as a function of $r$ for $Q=2$ with $L_h=2048$. }
\label{TableQ2}
\end{table}

\begin{table}[h]
\centering
\begin{tabular}{ | c || c | c | c | c || c || c ||   } 
\hline
   $\alpha$       &    1.5             &       2             &       3           &       4        \\
\hline
  $\text{FK}$     &    0.7922 (7)  &    0.5943 (7)   &  0.3036 (6)  &  0.1436 (5)  \\
  $\text{S}$     &    0.7939 (6)   &    0.5944 (8)  &  0.302 (1)     &    0.1408 (11) \\
  $\frac{\left<\varepsilon^{\text{FK}}\right>_{q=e^{-2\pi \alpha}}}{\left<\varepsilon^{\text{FK}}\right>_{q=e^{-2\pi }}}$  &    0.78819     &    0.59223   &  0.3031           &   0.14241   \\
  \hline
\end{tabular}
\caption{$c^{\text{X}}(q(\alpha))/c^{\text{X}}(q(1))$ as a function of $\alpha$ for $Q=3$ with $L_h=2048$. }
\label{TableQ3}
\end{table}

We compare now on the basis of Eq.~(\ref{CFTcoeff}) the value $c^{\text{S}}(q=e^{-2\pi})$, i.e. for the square domain $L_h=L_v$, reported in Table \ref{Tablea}, to the value:
\begin{equation}
\label{ansatz}
(2\pi)^{2\Delta_{1,2}}\left(C^{c\leq 1}\right)_{(1/2,0)(1/2,0)}^{(1,2)}\left<\varepsilon^{\text{S}}\right>=(2\pi)^{2\Delta_{1,2}}\left(C^{c\leq 1}\right)_{(1/2,0)(1/2,0)}^{(1,2)}\left<\varepsilon^{\text{FK}}\right>
\end{equation} 
for $Q\in [1,4]$. In the above identity we use the Eq.~(\ref{entor}). 
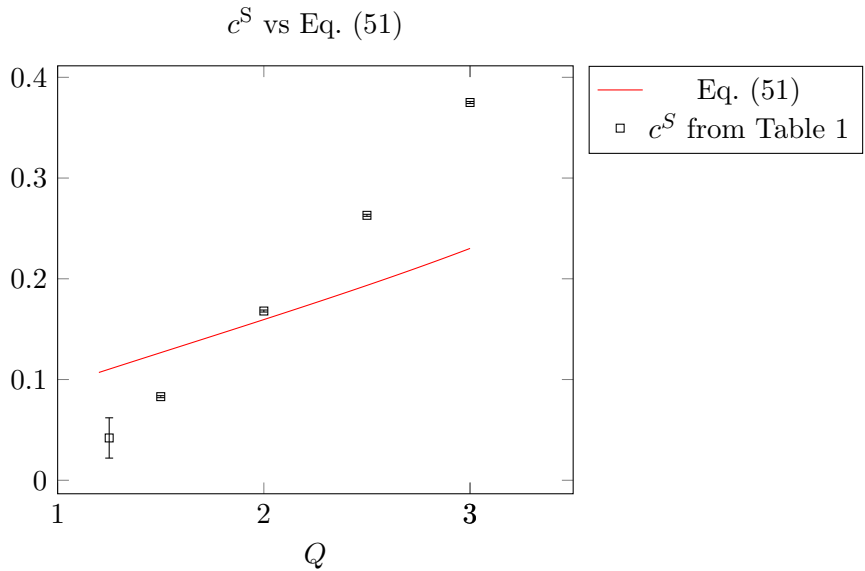
\begin{figure}
\begin{center}
\begin{tikzpicture}[scale=1]
\begin{scope}
\begin{axis}[
title={$c^{\text{S}}$ vs Eq.~(\ref{ansatz})},
	xlabel={$Q$},
	 xtick={1.0,2.0,3.0,3},
	xmin=1,
  xmax=3.5,
  legend pos=outer north east
	]
\addplot[red, smooth,mark= none] 
coordinates{
(1.2, 0.106998)(1.25, 0.110298)(1.3, 0.113589)(1.35, 0.116872)(1.4, 0.120149)(1.45, 0.123421)(1.5, 0.12669)(1.55, 0.129958)(1.6, 0.133226)(1.65, 0.136494)(1.7, 0.139765)(1.75, 0.14304)(1.8, 0.146319)(1.85, 0.149605)(1.9, 0.152898)(1.95, 0.156199)(2., 0.15951)(2.05, 0.162833)(2.1, 0.166167)(2.15, 0.169515)(2.2, 0.172879)(2.25, 0.176258)(2.3, 0.179656)(2.35, 0.183073)(2.4, 0.186511)(2.45, 0.189971)(2.5, 0.193456)(2.55, 0.196968)(2.6, 0.200507)(2.65, 0.204077)(2.7, 0.20768)(2.75, 0.211318)(2.8, 0.214994)(2.85, 0.218711)(2.9, 0.222473)(2.95, 0.226282)(3., 0.230142)};

\addplot[only marks,mark= square, mark size=1.5pt, fill=blue, error bars/.cd, y dir=both, y explicit ] 
coordinates{
(1.25,0.042) +- (0.0,0.02) (1.5,0.083) +- (0.0,0.001) (2,0.168) +- (0.0,0.001) (2.5,0.263) +- (0.0,0.001) (3,0.375) +- (0.0,0.001) };
%
\legend{Eq.~(\ref{ansatz}), $c^{S}$ from Table \ref{Tablea}}
\end{axis}
\end{scope}
\end{tikzpicture}
\caption{We compare the analytic value obtained from Eq.~(\ref{ansatz}) to the numerical values of $c^{\text{S}}$ from Table \ref{Tablea}.}
\label{ansatzvsc}
\end{center}
\end{figure}
From this comparison, shown in Fig.~(\ref{ansatzvsc}), we can surely conclude that:
\begin{equation}
\label{CCC}
C_{\Delta^{\text{S}}_{\text{conn}},\Delta^{\text{S}}_{\text{conn}}}^{\Delta^{\text{S}}_{\varepsilon}}\neq 0\neq \left(C^{c\leq 1}\right)_{(1/2,0)(1/2,0)}^{(1,2)}, \quad Q\in [1,4], Q\neq 2.
\end{equation}
For $Q=2$ instead, where the Eq.~(\ref{ansatz}) produces the value $0.15951$, the agreement with the numerical value $c^{S}=0.168$ is quite good.   In term of the structure constant,  one has to compare our  numerical finding $C_{\Delta^{\text{S}}_{\text{conn}},\Delta^{\text{S}}_{\text{conn}}}^{\Delta^{\text{S}}_{\varepsilon}}=0.17(1)$ with the Eq.~(\ref{3ensq2}), that predicts $C_{\Delta^{\text{S}}_{\text{conn}},\Delta^{\text{S}}_{\text{conn}}}^{\Delta^{\text{S}}_{\varepsilon}}\sim 0.163144$. 

For $Q\neq 2$, we cannot identify $C_{\Delta^{\text{S}}_{\text{conn}},\Delta^{\text{S}}_{\text{conn}}}^{\Delta^{\text{S}}_{\varepsilon}}$ to any known structure constant: they hint therefore at the existence of new CFT families.  The point $Q=3$ is useful to understand the degree of complexity of such theories. Differently from the Ising spin interfaces (described by the critical $O(1)$ loop models), the $3$-Potts spin interfaces can branch in a point separating the three colors. This is at the origin of the fact that these interfaces are described by generalizations of the $O(n)$ loops models, see \cite{Jacobsen2021} and references therein. In particular, while the $O(n)$ models are associated to the Lie ${\it sl}_2$ algebra (in turn related to the Virasoro algebras), the model describing the $Q=3$ spin interface is expected to be related to the ${\it sl}_3$ algebra ($W_3$  algebra).      
We report below our numerical value for the S-CFT structure constant in the $3-$Potts model~:
\beq
\label{Cdddq3}
C_{\Delta^{\text{S}}_{\text{conn}},\Delta^{\text{S}}_{\text{conn}}}^{\Delta^{\text{S}}_{\varepsilon}}=0.89 (1) \; \text{for} \; Q=3 \; .
\eeq
for future comparison with bootstrap solutions based on $W_3$ algebras, still to uncover.  
\subsection{The coefficent ${\bf c_T^{\text{X}}}$: the stress-energy tensor and the partition function}
In a domain of shape Eq.~(\ref{alpha}), we consider the coefficient $c_T^{\text{X}}$ related to the stress-energy tensor, see the Eq.~(\ref{CFTcoeff}). 
This coefficient vanishes for $\alpha=1$, i.e. in the case of a square domain $L_h=L_v$. As the stress-energy tensor contribution is a sub-sub-leading 
term, we use the difference as in Eq.~(\ref{diff2p}) where the sub-leading contributions cancel. A simple fit of our data, for FK clusters and S clusters, 
shows that the numerical result is proportional to $(r/L_v)^2$ with a great precision (the deviation of the power from two is a fraction of a percent). 
This high precision points out that higher order corrections from fields with spins are negligible.
 
\begin{figure}[!ht]
\begin{center}
\includegraphics[width=7.5cm]{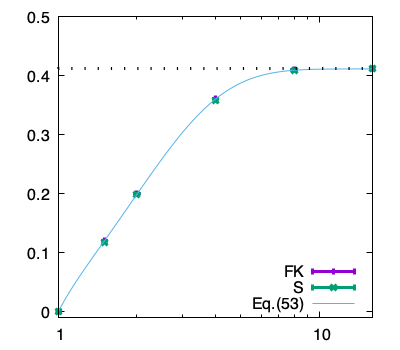}
\includegraphics[width=7.5cm]{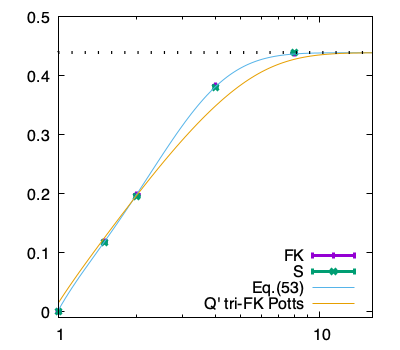}
\caption{\small 
$c_T^{\text{FK}}(q)$ and $(\Delta_{\text{conn}}^{\text{FK}}/\Delta_{\text{conn}}^{\text{SC}}) c_T^{\text{SC}}(q)$ as a function of $\alpha$. The measured data, for $L_v=512$,
is shown as points with error bars and the lines are given by Eq.(\ref{predict}). The horizontal dotted line is the exact result $2 \Delta_\text{conn}^{\text{FK}} \pi^2/3$ for the cylinder 
corresponding to the limit $\alpha\rightarrow \infty$ or $q\rightarrow 0$. 
Left panel, data for the Ising model and right panel, data for the $Q=3$ Potts model. This last figure contains also the prediction Eq.~(\ref{predict}) for the tri-critical FK clusters in the $Q' = (3+ \sqrt{5})/2$ Potts.
}
\label{PFit512}
\end{center}
\end{figure}

Next we fit the proportionality constant $4 a^{\text{X}}_0 c_{T}^{\text{X}}$. In Fig.~(\ref{PFit512}), we compare, for various $\alpha$, 
the obtained results (divided by the normalisation $4 a^{\text{X}}_0$) to the predictions of Eq.~(\ref{CFTcoeff}):
\begin{equation}
\label{predict}
c_{T}^{\text{X}}= \frac{2\Delta^{\text{X}}_{\text{conn}}}{c} \left< T^{\text{X}}\right>= \frac{2\Delta^{\text{X}}_{\text{conn}}}{c} \left(-2\pi^2 \partial_q \log{Z_{Q}}\right),
\end{equation} 
where we have used the Eq.~(\ref{ZQ}). The partition function $Z_{Q}$  is computed by truncating the $q$-series expansion given in \cite{FraSaZu87} to a sufficiently high order in $q$.  
For large values of $\alpha$, i.e. for $q\to 0$, one has $Z_{Q}\to q^{c/12}+o\left(q^{c/12}\right)$. This implies: 
\begin{equation}
\lim_{\substack{q\to 0\\(\alpha\to \infty)}}c_{T}^{\text{X}}=\lim_{q\to 0}\frac{-4\pi^2\Delta^{\text{X}}_{\text{conn}}}{c} \left(\partial_q \log{Z_{Q}}\right)=\frac{2 \Delta_\text{conn}^{\text{X}} \pi^2}{3}
\; .
\end{equation}
In Fig. (\ref{PFit512}) we show that the data converge towards the value $2 \Delta_\text{conn}^{\text{FK}} \pi^2/3$, represented by a horizontal dashed line. 
More interesting, we also show the data obtained for the S clusters but multiplied by $\Delta_{\text{conn}}^{\text{FK}}/\Delta_{\text{conn}}^{\text{S}}$. 
The measured data for the S clusters collapse perfectly, after this multiplication, on the data for the FK clusters. This confirms the validity of the 
Eq.~(\ref{predperco}), also taking into account that $Z^{\text{S}}=Z^{\text{1-tri-FK}}=Z_{Q=2}$.
 The data for $Q=3$ Potts model confirm again the validity of the Eq.~(\ref{predperco}) and the consistency with the Eq.~(\ref{ZQ}). Note that in this 
 latter case, $Z^{S}=Z_{Q}\neq Z^{\text{$Q'$-tri-FK}}$. The partition function of the tri-critical $Q'$-Potts model, of the same central charge as the 
 critical $Q$-Potts, does not appear in the $S$ clusters observables. 
\section{Wrapping for the $Q=3$ Potts model.}
In this section, we want to give a better description of the FK and S fixed points for the $Q=3$ Potts model, as shown in Fig.~(\ref{fig:phdiagr}). Our aim is to measure the quantities $y_{p_B}^{\text{S}}, y_J^{\text{S}}$ at the S fixed point. 

We are going to consider, both for the FK fixed point and the S fixed point, the probability of having 
a wrapping cluster\footnote{Other adimensional quantities like the Binder cumulant or $\xi/L$, with $\xi$ the second 
moment correlation length, would give similar results. The corrections to the scaling are larger for these two quantities 
compared to the wrapping probability that we employed in this work}.
This quantity was already considered in the past. In particular, in \cite{Fortunato2002} it was observed that the probability 
of having a percolating Ising S cluster scales as $(J-J_c)L^{1/\nu}$ with $\nu = 2 - 1/\Delta^{S}_\epsilon=1$, the correlation length exponent of the Ising model 
\footnote{In \cite{Fortunato2002}, crossing clusters were considered with free periodic conditions on the lattice. Wrapping clusters with periodic boundary conditions 
give better results since there are less finite-size corrections\cite{Newman2001}.}.
Similar results were also obtained for the $Q=3$ Potts model with $\nu \simeq 5/6$, the correlation length exponent of the critical $Q=3$ Potts model. In \cite{Fortunato2002}, the S clusters were considered with $p_B=1$.
\begin{figure}[!ht]
\begin{center}
\includegraphics[width=12cm]{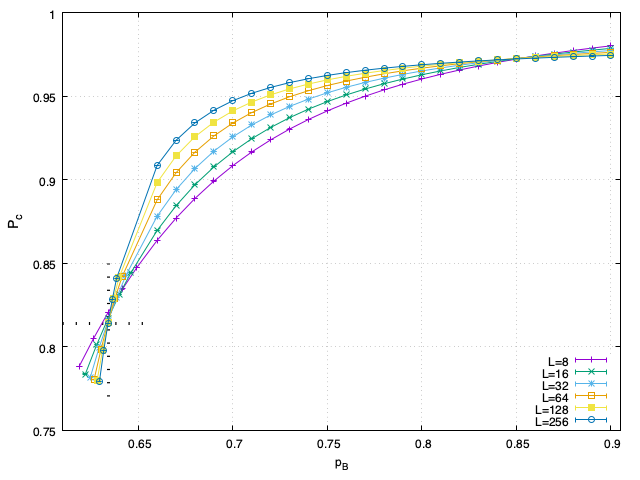}
\caption{\small 
Crossing points for ${\bf{P_c}} (J_c,p_B,L)$ as a function of $p_B$ for the values of $L$'s shown in the caption.
}
\label{FigACQ3}
\end{center}
\end{figure}
For the $Q=3$ Potts model, it is expected that the S fixed point corresponds to \cite{Delfino13}
\beq
\label{SCQ3}
p_B^{\text{S}} \simeq 1 - e^{-1.7 J_c}  \; .
\eeq
The coefficient $1.7$ was obtained by looking the magnetic effective exponent and determining 
$p_B^{\text{S}}$ such that this exponent is close to the expected result without much corrections. We will try to do better here. 

In the following, we are going to consider the probability that a $p_B$ cluster (see definition in Section \ref{secFK})  is wrapping around a toric $L\times L$ lattice.
We denote by ${\bf{P_c}} (J,p_B,L)$ this probability\footnote{It can be a cluster wrapping in the horizontal direction or the vertical direction or in both directions.}, with $J$ the inverse temperature at which we equilibrate the system.
In Fig.~\ref{FigACQ3}, we show the results of a numerical measurement of ${\bf{P_c}} (J_c,p_B,L)$ as a function of $p_B$ and $L$. 
We observe that the data is crossing for two values of $p_B$ which corresponds to two fixed points~:
a first one for FK clusters at $p_B^{\text{FK}}=1-e^{-J_c}=0.63397$ and with a value ${\bf{P_c}} (J_c,p_B^{\text{FK}})=0.813667$ in the large size limit \cite{Arguin2001}.
These two exact values are shown as dotted lines. 
A second one is the S fixed point,  for $p_B^{\text{S}} \simeq 0.84$ and ${\bf{P_c}} (J_c,p_B^{\text{S}})\simeq 0.97$.
Our aim here is to determine the operators corresponding to the perturbations of the S fixed point. 
But first, we will show how it works for the FK fixed point.

\subsection{FK fixed point}
We consider the probability ${\bf{P_c}}$ that a FK cluster is wrapping around the lattice which is known exactly  for the critical $Q$ Potts models \cite{Arguin2001}. 
On a finite size lattice and close to the FK fixed point, this quantity behaves as
\beq
{\bf{P_c}} (J,p_B,L) = f((J-J_c) L^{y_J^{\text{FK}}},(p_B-p_B^{\text{FK}})L^{y_{p_B}^{\text{FK}}})(1  + \alpha_\omega L^{-\omega} + \cdots )\; ,
\eeq
where the $y_J^{\text{FK}}$ and  $y_{p_B}^{\text{FK}}$ are expected to correspond respectively to  $y_J^{\text{FK}}=6/5$ and $y_{p_B}^{\text{FK}}=7/20$, as obtained by setting $\beta^2=5/6$ in the Eq.~(\ref{FKexp}). The  $p_B^{\text{FK}}$ is defined in the Eq.(\ref{FKSC}) and the $\omega$ is the leading correction to the scaling  associated to the sub leading thermal operator \cite{Nienhuis82}. Its value is given by $\omega= \Delta_{(1,3)}-2 = 0.8$. 

We first consider the case of a perturbation of $p_B$ close to $p^{\text{FK}}_{B}$ while keeping $J=J_c$~:
\bea 
\label{Pcfk}
{\bf{P_c}} (J_c,p_B,L) &=& f(0,(p_B-p_B^{\text{FK}})L^{y_{p_B}^{\text{FK}}}) (1 + \alpha_\omega L^{-\omega} + \cdots ) \\
  &=&( f(0)  + \alpha_1 (p_B-p_B^{\text{FK}})L^{y_{p_B}^{\text{FK}}} + \cdots)(1 +  \alpha_\omega L^{-\omega} + \cdots  ) \; . \nn
\eea
In order to determine $p_B^{\text{FK}}, y_p^{FK}$ and $\omega$, we will use the quotients-method approach to finite-size scaling\cite{Victor}.
We compute the wrapping for a pair of lattices sizes $L$ and $2 L$ and determine the critical value $p_B^c(L)$ such that  
\beq
\label{cross}
{\bf{P_c}} (J_c,p_B^c(L),L)    =  {\bf{P_c}} (J,p_B^c(L),2 L) \; ,
\eeq
which corresponds to 
\beq
\alpha_1 (p_B^c-p_B^{\text{FK}})L^{y_{p_B}^{\text{FK}}}  + \alpha_\omega L^{-\omega} = \alpha_1 (p_B^c-p_B^{\text{FK}})(2 L)^{y_{p_B}^{\text{FK}}} + \alpha_\omega (2 L)^{-\omega} \; .
\eeq
From this, we determine~:
\beq
\label{crossq31}
p_B^c(L) = p_B^{\text{FK}} + \alpha L^{-\omega-y_{p_B}^{\text{FK}}}\; ,
\eeq
with $\alpha=\frac{\alpha_\omega}{\alpha_1}(1-2^{-\omega})/(2^{y_{p_B}^{\text{FK}}}-1)$ a constant and then one can determine numerically $\omega+y_{p_B}^{\text{FK}}$. With the data shown in Fig.~(\ref{FigACQ3}) and keeping only the data for $L \geq 32$,
we measure $\omega+y_{p_B}^{\text{FK}} = 1.24 (3)$. 
Next, by using the previous relation in the expansion of ${\bf{P_c}} (J_c, p_B^c(L),L)$, one gets
\beq
\label{crossq32}
{\bf{P_c}} (J_c, p_B^c(L),L) = f(0) +  (\alpha+\alpha_\omega)L^{-\omega} + \cdots
\eeq
 which gives a direct way to measure $\omega$. Using the exact result $f(0) = 0.813667$ \cite{Arguin2001}, we measure
 $\omega = 0.85 (2)$ which is close to the expected value $\omega =0.8$. And we obtain $y_{p_B}^{\text{FK}} \simeq 0.39$ which is 
 also close to the exact result $y_{p_B}^{\text{FK}}=7/20=0.35$. 
\begin{figure}[!ht]
\begin{center}
\includegraphics[width=10cm]{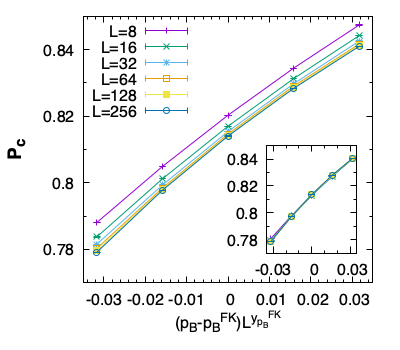}
\caption{\small 
${\bf{P_c}}(\beta_c,p_B,L)$ vs $(p_B-p_B^{\text{FK}}) L^{y_{p_B}^{\text{FK}}}$. In the inset, we remove a $L^{-\omega}$ contribution. 
}
\label{FigCQ3b}
\end{center}
\end{figure}
Last, to check the consistency of our results, we show in Fig.~(\ref{FigCQ3b}), ${\cal P}_c$ vs. $(p_B-p_B^{\text{FK}}) L^{y_{p_B}^{\text{FK}}}$ on the left panel, 
with $y_{p_B}^{\text{FK}} = 7/20$. 
There is a scaling but with a shift which depends on the linear size $L$. In the right panel, we subtract a term $\simeq L^{-\omega}$  
and get a nice collapse of the data. Note also that we observe a curvature in the data of Figs.~(\ref{FigACQ3})-(\ref{FigCQ3b}) 
indicating the existence of highest order corrections terms which we have not tken into account. 

We next consider a thermal perturbation. Thus, one has 
\bea 
{\bf{P_c}} (J,p_B^{\text{FK}},L) &=& f((J-J_c)L^{y_J^{\text{FK}}},0)(1 + \alpha_\omega L^{-\omega}  + \cdots) \\
  &=& (f(0)  + \rho_1 (J-J_c)L^{y_J^{\text{FK}}} + \cdots ) ( 1 + \alpha_\omega L^{-\omega} + \cdots ) \; . \nn
\eea
By measuring the wrapping of the data for increasing sizes, one can determine $y^{\text{FK}}_J$ and $\omega$ for which one obtains $y_J^{\text{FK}}=1.2=6/5$. 
Note that in that case the measurement is much simple since $y_J^{\text{FK}} + \omega = 2$ which is large and then the wrapping is converging fast. 

\subsection{S fixed point}
\begin{figure}[!ht]
\begin{center}
\includegraphics[width=7.4cm]{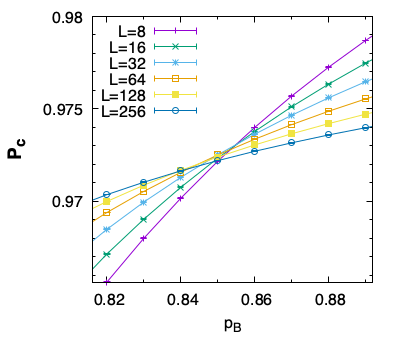}
\includegraphics[width=7.4cm]{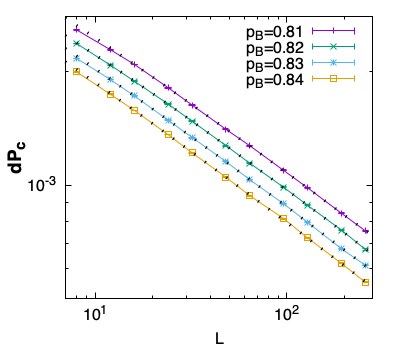}
\caption{\small 
Left panel : Crossing points for S clusters as a function of $p_B$~: ${\bf{P_c}} (J_c,p_B,L)$ vs $p_B$ for the values of $L$'s shown in the caption. 
Right panel : ${\bf{P_c}}(J,p_B+dp_B,L)-{\bf{P_c}}(J,p_B,L)$ vs. $L$ for values of $p_B$ shown in the caption and $dp_B=0.01$.
}
\label{FigSC3}
\end{center}
\end{figure}
Next we consider the fixed point associated to the S clusters. The wrapping probability is expect to have the following form
\beq
\label{Pcsc}
{\bf{P_c}} (J,p_B,L) = g((J-J_c)L^{y_J^{\text{S}}},(p_B-p_B^{\text{S}})L^{y_{p_B}^{\text{S}}})( 1+ \alpha_\omega' L^{-\omega'} + \cdots)  \\
\eeq
which is similar to the Eq.~(\ref{Pcfk}). The exact value for $y_{p_B}^{\text{S}}$ has been conjectured to be given by the Eq.~(\ref{ypb}). 
In the left part of Fig.~(\ref{FigSC3}) we show the crossing of ${\bf{P_c}} (J_c,p_B,L)$  as a function of $p_B$ close to the value $p_B^{\text{S}} \simeq 0.84$. 
Moreover, close to this fixed point, the slope of ${\bf{P_c}} (J_c,p_B,L)$ as a function of $p_B$ decreases as on increases $L$.
This is in agreement with the fact that the fixed point is attractive and $y_{p_B}^{\text{S}} < 0$. 

A direct measurement of $y_{p_B}^{\text{S}}$ can be obtained by considering 
the difference of wrapping probabilities for two close values of $p_B$. Indeed, from the Eq.~(\ref{Pcsc}) we expect
\beq
\label{DifCorSC}
{\bf{P_c}} (J_c,p_B+dp_B,L) -{\bf{P_c}} (J_c,p_B,L)  = \rho_1  dp_B L^{y_{p_B}^{\text{S}}}  + \cdots \; . 
\eeq
In the right part of Fig.~(\ref{FigSC3}) we show the measurement of ${\bf{P_c}}(J,p_B+dp,L)-{\bf{P_c}}(J,p_B,L)$ for $p_B=0.81, \cdots, 0.84$ and $dp=0.01$. 
The behaviour is similar for all these choices and the differences are due to largest order corrections. A fit to the form ~(\ref{DifCorSC}), including a second order correction term 
$\alpha_2 L^{2 y_{p_B}^{\text{S}}}$, gives $-y_{p_B}^{\text{S}} = 0.389 (5) ;  0.401 (6) ;  0.396 (7) ;  0.395 (9)$. 
This can be compared with the value expected for the $\Delta^{\text{tri-FK}}_{\text{piv}}$ pivotal fields of the tricritcal 
Potts model, $\Delta^{\text{tri-FK}}_{\text{piv}}=\Delta_{0,2} =2.38333$ at $Q=3$, yielding $y_{p_B}^{\text{S}}=-0.38333$. The measured values  are close and compatible with this prediction.

\begin{figure}[!ht]
\begin{center}
\includegraphics[width=10cm]{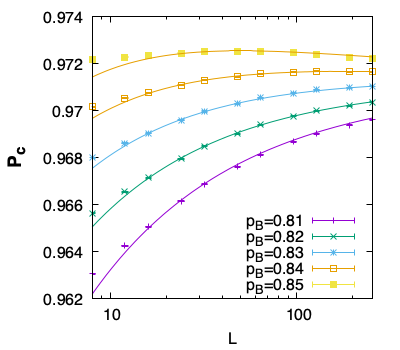}
\caption{\small 
${\bf{P_c}} (\beta_c,p_B,L)$ vs $L$ for the values of $p_B$ shown in the caption. The lines are a fit to the form eq.~(\ref{Pcl2}), see the text for details. 
}
\label{AFit}
\end{center}
\end{figure}

We will now try a more general fit of the data by considering the following expansion of the Eq.~(\ref{Pcsc})~:
\beq
\label{Pcl2}
{\bf{P_c}} (J_c,p_B,L) =   g(0)  + \alpha_1 (p_B-p_B^{\text{S}})L^{y_{p_B}^{\text{S}}}  + \alpha_2 (p_B-p_B^{\text{S}})^2 L^{2 y_{p_B}^{\text{S}}} + g(0) \alpha_\omega' L^{-\omega'} \; . 
\eeq
In Fig.~(\ref{AFit}), we show data for $p_B=0.81, \cdots, 0.85$. Then, a fit is done with the data $p_B \in [0.82,0.84]$ and $L \geq 16$. 
A first fit with free parameters gives $y_{p_B}^{\text{S}} = -0.385 (7)$ and $\omega'=0.65$. 
We then repeat the fit while imposing $y_{p_B}^{\text{S}} = -0.38333$. The result is shown in Fig.~(\ref{AFit}), in which the continuous lines correspond to the best fit. 
Note that even if the fit is done for  $p_B \in [0.82,0.84]$, the agreement is still good for $p_B=0.81$ and $p_B=0.85$. 
The best fit gives the following values
\bea
\label{resfit}
g(0) = 0.9712 (1) \; ; \; \alpha_1 = 0.55 (1) \; &;& \; \alpha_2 = -9 (2) \; ; \; g(0) \alpha_\omega' = -0.018 (1) \nn \\
\omega'= 0.64 (4) \; &;& \; p_B^{\text{S}} = 0.825 (2)  \; .
\eea
We also made attempts with adding more corrections without improvement or change in the main results. Last, note that 
in the Eq.~(\ref{Pcl2}) expansion, there are three powers with close dimension, $y_{p_B}^{\text{S}}, -\omega', 2 y_{p_B}^{\text{S}} = -0.38333, -0.64 , -0.76666$. A consequence 
is that the obtained values of $p_B^{\text{S}}$ and $\omega'$ are strongly correlated and also the errors on these quantities.

\begin{figure}[!ht]
\begin{center}
\includegraphics[width=7cm,height=6cm]{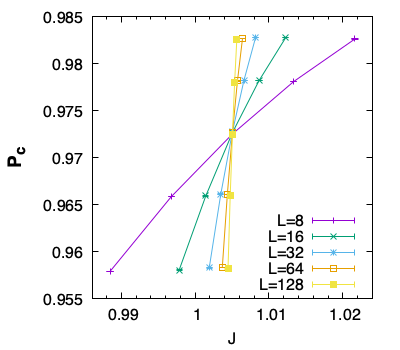}
\includegraphics[width=7cm,height=6cm]{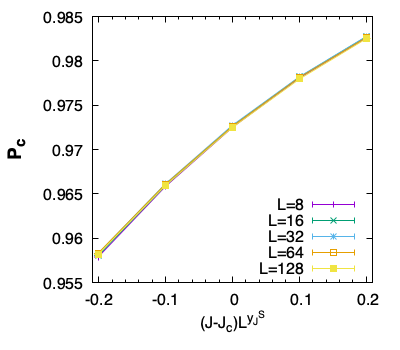}
\caption{\small 
Left panel :  ${\bf{P_c}} (J,p_B^{\text{S}},L)$ vs. $J$. Right panel : ${\bf{P_c}} (J,p_B^{\text{S}},L)$ vs $(J-J_c) L^{y_J^{\text{S}}}$.}
\label{FigCQ3d}
\end{center}
\end{figure}

Last, we are interested in determining the exponent $y_J^{S}$ defined by 
\beq 
{\bf{P_c}} (J,p_B,L) = g((J-J_c)L^{y_J^{\text{S}}},(p_B-p_B^{\text{S}})L^{y_{p_B}^{\text{S}}})(1 + \alpha_\omega' L^{-\omega'} )  \; .
\eeq
We can develop in function of the two variables
\beq
{\bf{P_c}} (J,p_B,L) = g(0) +  a_1(J-J_c)L^{y_J^{\text{S}}} + \alpha_1 (p_B-p_B^{\text{S}})L^{y_{p_B}^{\text{S}}}+ g(0) \alpha_\omega' L^{-\omega'}   \; .
\eeq
But then, since $y_{p_B}^{\text{S}} < 0$, one expects that this will be dominated by the correction $(J-J_c)L^{y_J^{\text{S}}}$. 
In the left panel of Fig.~(\ref{FigCQ3d}), we show the result of the measurement of ${\bf{P_c}}  (J,p_B,L)$ for  
$p_B =1 -e^{-J}   \simeq 1-e^{-1.9 J_c} \simeq p_B^{\text{S}}$  as a 
function of $J$ for $J \simeq J_c$. We observe a crossing of the data at $J_c$. On the right panel,
we show the same quantity but as a function of the scaling variable $(J-J_c)L^{y_J^{\text{S}}}$. We obtain a perfect scaling 
with 
\beq
\label{Finaly}
y_J^{\text{S}}=1.20 (1) \; . 
\eeq
This is of course the result already obtained long time ago in \cite{Fortunato2002}. 
%

In summary, we obtained the following numerical results
\bea
p_B^{\text{S}} = 0.825 (2)  \; &;& \;  {\bf{P_c}} (J_c,p_B^{\text{S}},L\rightarrow \infty) = 0.9712 (1)  \nn \\
\omega'= 0.64 (4) \; &;& \;  y_J^{\text{S}}=1.20 (1) \; . 
\eea

\section{Conclusions}
We have considered clusters of like spin, the S clusters, in the $Q$-Potts model. Using  Monte Carlo simulations,
we studied the S clusters on a toric $L_h\times L_v$ square lattice for values of $Q\in [1,4]$. The main motivation behind this work was to investigate the existence of a CFT, the S-CFT, that describes the connectivity of the critical S clusters. We measured the universal finite size corrections of the two-point connectivity $p^{\text{S}}_{12}$ and we compare the data to the CFT predictions in the Eq.~(\ref{predperco}). We could probe the dominant corrections and the sub-dominant ones. These latters, which depend on the orientation of the two points and are not vanishing for rectangular domains $L_h\neq L_v$, probe the stress-energy tensor and the  partition function of the theory.  The Monte Carlo data are perfectly compatible with the Eq.~(\ref{predperco}) and with all the established results on the S fixed point. Our result support therefore the existence of a consistent S-CFT which remains at the present unknown. 

We provided new insights on the energy fields $\varepsilon^{\text{S}}$ of the S-CFT theory.  In particular we determine its conformal dimension $\Delta_{\varepsilon}^{\text{S}}$, see the Eq.(\ref{yj}), its one-point correlation function, see the Eq.~(\ref{entor}) and its three-point function with the connectivity field, see the Eqs~ (\ref{3ens})-(\ref{3ensq2}). For $Q=2$, where the S fixed point is identified with the tri-critical $1$-Potts point, our numerical evaluation for $C_{\Delta^{\text{S}}_{\text{conn}},\Delta^{\text{S}}_{\text{conn}}}^{\Delta^{\text{S}}_{\varepsilon}}$ is in good agreement with the corresponding imaginary Liouville structure constants, see the Eq.~(\ref{3ensq2}). We find this result consistent with the known results for the connectivity fields Eq.~(\ref{3-S}), proved in \cite{Ang21} for the CLE$_3$ loop model. For $Q\neq 2$ instead, the $C_{\Delta^{\text{S}}_{\text{conn}},\Delta^{\text{S}}_{\text{conn}}}^{\Delta^{\text{S}}_{\varepsilon}}$ does not correspond to any known bootstrap solutions. In \cite{Jacobsen2021} a generalization of the $O(n)$ loops models, which describes the boundaries of the S clusters for $Q=3$, have been introduced. The results presented here, see for instance the Eq.~(\ref{Cdddq3}), should be relevant for the CFT associated to this family of integrable models.  

The validity of our analysis is backed up by the measurements of the S-cluster wrapping probability for $Q=3$, which provided an independent evaluation of $\Delta_{\varepsilon}^{\text{S}}$, see Eq.~(\ref{Finaly}). We determined for the first time the RG exponent of the irrelevant perturbation as well as the correction to the scaling at the S fixed point, see Fig.~(\ref{AFit}). Finally we mention that a new, more compact, expression for the torus one-point function of the $Q$-Potts energy field has been given in the Eq.~(\ref{resulte}).

\vspace{0.25cm}

\noindent
{\bf Acknowledgements}
We are grateful to Sylvain Ribault, Rongvoram Nivesvivat and Xin Sun  for helpful discussions. We thank in particular Gesualdo Delfino, Nina Javersat and Jacopo Viti for comments on the draft of this article. We would like to thank the organisers and participants of Bootstat 2021 where this work was partially done.

\vspace{0.25cm}

\appendix

\section{Computation of $\left<\varepsilon^{FK}\right>$}
\label{efktorus}
The  $\left<\varepsilon^{FK}\right>$ was computed in \cite{Zuber89} using a Coulomb gas approach. In particular, an expression in terms of an integral over the $1-$ cylcle of the torus was given, see the Eq.~(30) in \cite{Zuber89}.
Here we determine completely the $\left<\varepsilon^{FK}\right>$ in a more explicit form by using the Poghossian identity, see \cite{Hadasz2010}, relating the one-point torus conformal block to a four-point conformal block on the sphere.

The field $\varepsilon^{FK}$ is associated to a degenerate representation $\mathcal{R}_{\Delta_{1,2}}$ with a vanishing null state at level two. This properties fixes the fusion rules of this field:
\begin{equation}
\label{fus2lev}
\varepsilon^{FK}\otimes V_{r,s} = V_{r,s+1}\oplus V_{r,s-1}.
\end{equation} 

The one-point torus conformal block $ \left<V_{1,2} \right>$, on which $\left<\varepsilon^{FK}\right>$ is built, is represented by the diagram: 
\begin{center}
\begin{tikzpicture}[baseline=(current  bounding  box.center), very thick, scale = .4]
\draw (0,0)  -- (0,-3);
\draw (0,-1.5) node[right]{$ V_{1,2}$};
\draw (0,-5) circle(1.9cm);
\draw (0,-7) node [below] {$V_{(0,\frac12)}$};
\end{tikzpicture}
\end{center}
where only the internal channel $V_{(0,\frac12)}=V_{(0,-\frac12)}$ is allowed by the fusion (\ref{fus2lev}).  
 It is convenient here to use the charge $\alpha$ representation for the conformal dimension:
\begin{equation}
\Delta_{\alpha} = \alpha\left(\alpha -\beta+\frac{1}{\beta}\right),
\end{equation}
The Poghossian identity can be representated by this diagram:
\begin{equation}
\begin{tikzpicture}[baseline=(current  bounding  box.center), very thick, scale = .4]
\draw (-3,0)  -- (-3,-3);
\draw (-3,-1.5) node[right]{$ V_{\alpha}$};
\draw (-3,-5) circle(1.9cm);
\draw (-3,-7) node [below] {$V_{\beta}$};
\draw(-1,-5) node[right]{$\rightarrow$};
\draw (4,-5)--(5,-5);
\draw (5,-5)--(5,-4);
\draw (5,-5)--(8,-5);
\draw (8,-5)--(8,-4);
\draw (7,-5)--(9,-5);
\draw (4,-5) node[left]{$V_{\alpha_{\infty}}(\infty)$};
\draw (5,-4) node[above]{$V_{\tilde{\tilde{\alpha}}}(1)$};
\draw (8,-4) node[above]{$V_{\tilde{\alpha}}(x)$};
\draw (9,-5) node[right]{$V_{\alpha_0} (0)$};
\draw (6.5,-5) node[below]{$V_{\beta}$};
\end{tikzpicture}
\end{equation}
with:
\begin{align}
\tilde{\alpha} &=\frac{\beta}{4} + \frac{\alpha}{2}, \quad \tilde{\tilde{\alpha}}=\frac{\beta}{4} -\frac{1}{2\beta}+\frac{\alpha}{2}\\
\alpha_0 &=\alpha_{\infty} = \frac{\beta}{4} -\frac{1}{2\beta}\\
x &= \left(\frac{\theta_2(q)}{\theta_3(q)}\right)^4
\end{align}
Taking into account the conversion between the  $(b, \lambda)$ parameter  used in \cite{Hadasz2010} and the ones $(\beta, \alpha)$ used here ,  $b\to -i \beta^{-1}$, and $\lambda\to i\alpha = i(\beta-\beta^{-1})+i\lambda/2$, the above equations correspond to the second of the three forms of the Pogosshian identity displayed in the Eq.~(13) of \cite{Hadasz2010}. We refer the reader to \cite{Hadasz2010} where this identity is explained in full detail.

One can easily verify that:
\begin{equation}
\tilde{\alpha}+\tilde{\tilde{\alpha}}+\alpha_0+\alpha_{\infty}=-\frac{1}{\beta}+\beta, \quad \text{for}\; \alpha=\frac{1}{2\beta},
\end{equation}
that means that the external charges associated to the one point function $\left<V_{(1,2)}\right>_q$, satisfies a neutrality Coulomb gas condition with no screenings.  This implies that the four-point conformal block, with internal channel $\alpha_{int}=\alpha_0+\tilde{\alpha}=-\beta^{-1}/2+\beta/4 $, which corresponds to $V_{\alpha_{int}}=V_{0,\frac12}$, takes a very simple form:
\begin{equation}
\begin{tikzpicture}[baseline=(current  bounding  box.center), very thick, scale = .4]
\draw (4,-5)--(5,-5);
\draw (5,-5)--(5,-4);
\draw (5,-5)--(8,-5);
\draw (8,-5)--(8,-4);
\draw (7,-5)--(9,-5);
\draw (4,-5) node[left]{$V_{\alpha_{\infty}}(\infty)$};
\draw (3.5,-4) node[above]{$V_{\frac{1}{4}\left(-\beta+\frac{1}{\beta}\right)}(1)$};
\draw (9,-4) node[above]{$V_{\frac{1}{4}\left(-\beta -\frac{1}{\beta}\right)}(x)$};
\draw (9,-5) node[right]{$V_{\alpha_0} (0)$};
\draw (6.5,-5) node[below]{$V_{\frac{\beta}{2}-\frac{1}{4\beta}}$};
\end{tikzpicture}
=x^{-2\alpha_0 \tilde{\alpha}}(1-x)^{-2 \tilde{\tilde{\alpha}}\tilde{\alpha}} = x^{\frac{(-2 + \beta^2) (1 + \beta^2)}{8 \beta^2}}(1-x)^{\frac{-1 + \beta^4}{8 \beta^2}}
\end{equation}
Collecting all the above results and the definition of $ \left<\varepsilon^{FK}\right>$, see  Eq.~(2.18) in \cite{Javerzat19}, we find:
\begin{align}
\label{resulte}
\left<\varepsilon^{FK}\right>&=\frac{(Q-1)}{Z_Q}\; \left(C^{c\leq 1}\right)_{(0,\frac12)(0,\frac12)}^{(1,2)}\;\times\nonumber \\
&\times \;\frac{|q|^{2\Delta_{1,2}-\frac{c-1}{12}}}{|\eta(q)|^2}  |x(q)|^{2\delta_0}|1-x(q)|^{2\delta_1}|\theta_{3}(q)|^{-2\delta_3}|16 q|^{-2\delta_0}\\
\delta_0&= \frac{1}{16}\left(-2 - \frac{5}{\beta^2} + 2 \beta^2\right), \; \delta_1=\frac{-3 + \beta^4}{8 \beta^2},\;\delta_3=-\frac{3}{2\beta^2}
\end{align}
Note that the factor $Q-1$ comes from the multiplicity of the $\mathcal{R}_{0,\frac12}$ representation in the Potts model.

\end{document}